\documentclass[10pt,letterpaper]{article}
\usepackage[letterpaper,margin=1in,marginparwidth=0.8in,marginparsep=4pt]{geometry}
\usepackage[T1]{fontenc}
\usepackage[utf8]{inputenc}
\usepackage{mathptmx}          %
\usepackage{microtype}
\usepackage{amsmath,amssymb,amsthm}
\usepackage{mathrsfs}          %
\usepackage{accents}           %
\usepackage{xcolor}
\usepackage[normalem]{ulem}    %
\usepackage{needspace}
\newenvironment{blackmath}{\color{black}}{}%
\usepackage{graphicx}
\usepackage{booktabs}
\usepackage{array}
\usepackage{multirow}
\usepackage{xspace}
\usepackage{enumitem}
\usepackage{caption}
\usepackage{subcaption}
\usepackage{algorithm}
\usepackage{placeins}
\usepackage{wrapfig}
\usepackage{float}             %
\usepackage[noend]{algpseudocode}
\usepackage{tikz}
\usetikzlibrary{positioning,arrows.meta,patterns,calc,fit,backgrounds}
\usepackage[numbers,sort&compress]{natbib}
\usepackage{xr}                %
\usepackage[hypertexnames=false,colorlinks=true,linkcolor=blue,citecolor=blue,urlcolor=blue]{hyperref}
\usepackage[capitalise,noabbrev]{cleveref}

\setlist{itemsep=1pt,topsep=2pt,leftmargin=1.2em}

\AtBeginDocument{\setlength{\abovedisplayskip}{4pt plus 1pt minus 1pt}\setlength{\belowdisplayskip}{4pt plus 1pt minus 1pt}\setlength{\abovedisplayshortskip}{2pt}\setlength{\belowdisplayshortskip}{3pt}}%
\crefname{equation}{Eq.}{Eqs.}\Crefname{equation}{Eq.}{Eqs.}
\crefname{algorithm}{Alg.}{Algs.}
\crefname{figure}{Fig.}{Figs.}\Crefname{figure}{Fig.}{Figs.}
\crefname{section}{Sec.}{Secs.}\Crefname{section}{Sec.}{Secs.}
\crefname{subsection}{Sec.}{Secs.}\Crefname{subsection}{Sec.}{Secs.}
\crefname{appendix}{App.}{Apps.}\Crefname{appendix}{App.}{Apps.}

\renewcommand{\cite}[1]{\citep{#1}}

\renewcommand{\paragraph}[1]{\vspace{2pt}\noindent\textbf{#1}\enspace}

\theoremstyle{plain}

\theoremstyle{definition}

\newif\ifdraft
\drafttrue

\definecolor{todocolor}{HTML}{C0392B}
\definecolor{boxcolor}{HTML}{7F8C8D}

\ifdraft
  \newcommand{\todo}[1]{}%
  \newcommand{\milan}[1]{{\color{blue!70!black}\textbf{[Milan:} #1\textbf{]}}}
  \newcommand{\budget}[1]{}%
  \newcommand{\placeholder}[2]{}%
\else
  \newcommand{\todo}[1]{}
  \newcommand{\milan}[1]{}
  \newcommand{\budget}[1]{}
  \newcommand{\placeholder}[2]{}
\fi
\newenvironment{edited}{}{}%
\definecolor{lhpink}{HTML}{D81B7A}%

\newcommand{\fixdel}[1]{}

\newcommand{\LN}{LinearKnot\xspace}            %
\newcommand{\BeamDP}{Beam-DP03\xspace}          %
\newcommand{\LRDP}{LR-DP03\xspace}              %
\newcommand{\DPthree}{DP03\xspace}              %
\newcommand{\DPnine}{DP09\xspace}               %
\newcommand{\NUPACK}{NUPACK\xspace}
\newcommand{\LinearPartition}{LinearPartition\xspace}
\newcommand{\LinearFold}{LinearFold\xspace}
\newcommand{\ThreshKnot}{ThreshKnot\xspace}
\newcommand{\IPknot}{IPknot\xspace}
\newcommand{\HotKnots}{HotKnots\xspace}
\newcommand{\PseudoBase}{PseudoBase\xspace}
\newcommand{\ArchiveII}{ArchiveII\xspace}
\newcommand{\nt}{\,nt\xspace}                   %

\renewcommand{\vec}[1]{\ensuremath{\boldsymbol{#1}}}
\newcommand{\vecx}{\ensuremath{\vec{x}}\xspace}        %
\newcommand{\vecy}{\ensuremath{\vec{y}}\xspace}        %
\newcommand{\Yset}{\ensuremath{\mathcal{Y}}}           %

\newcommand{\order}{\ensuremath{\mathcal{O}}}         %

\newcommand{\ind}[1]{\ensuremath{[\![#1]\!]}}          %
\newcommand{\pair}[2]{#1\cdot #2}                      %
\newcommand{\RT}{RT}
\newcommand{\Zfn}{\ensuremath{Z}}                      %
\newcommand{\Gfe}{\ensuremath{\Delta G^\circ}}               %

\newcommand{\betaone}{\ensuremath{\beta_1}}            %
\newcommand{\betaonem}{\ensuremath{\beta_1^{\mathrm{m}}}} %
\newcommand{\betaonep}{\ensuremath{\beta_1^{\mathrm{p}}}} %
\newcommand{\betatwo}{\ensuremath{\beta_2}}            %
\newcommand{\betathree}{\ensuremath{\beta_3}}          %
\newcommand{\omegaone}{\ensuremath{\omega_1}}          %
\newcommand{\omegaonem}{\ensuremath{\omega_1^{\mathrm{m}}}}
\newcommand{\omegaonep}{\ensuremath{\omega_1^{\mathrm{p}}}}
\newcommand{\alphaone}{\ensuremath{\alpha_1}}
\newcommand{\alphatwo}{\ensuremath{\alpha_2}}
\newcommand{\alphathree}{\ensuremath{\alpha_3}}
\newcommand{\Ein}{E^{\mathrm{in}}}                     %
\newcommand{\Eout}{E^{\mathrm{out}}}                   %
\newcommand{\Lmax}{\ensuremath{L}}                     %

\newcommand{\Qb}{Q^{\mathrm{b}}}       %
\newcommand{\Qm}{Q^{\mathrm{m}}}       %
\newcommand{\Qz}{Q^{\mathrm{z}}}       %
\newcommand{\Qp}{Q^{\mathrm{p}}}       %
\newcommand{\Qg}{Q^{\mathrm{g}}}       %
\newcommand{\Qgl}{Q^{\mathrm{gl}}}     %
\newcommand{\Qgr}{Q^{\mathrm{gr}}}     %
\newcommand{\Qgls}{Q^{\mathrm{gls}}}   %
\newcommand{\Qgrs}{Q^{\mathrm{grs}}}   %
\newcommand{\Nin}{\ensuremath{\mathcal{N}^{\mathrm{in}}}}   %
\newcommand{\Nout}{\ensuremath{\mathcal{N}^{\mathrm{out}}}} %
\newcommand{\wloop}[4]{\ensuremath{w(#1,#2\mid #3,#4)}}     %
\newcommand{\Fone}{\ensuremath{F_1}}   %
\newcommand{\Ftwo}{\ensuremath{F_2}}
\newcommand{\Fthree}{\ensuremath{F_3}}

\newcommand{\Zin}{Z^{\mathrm{in}}}
\newcommand{\Zout}{Z^{\mathrm{out}}}
\newcommand{\bone}{\ensuremath{b_1}}   %
\newcommand{\btwo}{\ensuremath{b_2}}   %
\newcommand{\bfour}{\ensuremath{b_4}}  %
\newcommand{\bV}{\ensuremath{b_V}}    %

\newcommand{\Cpref}{C}
\newcommand{\Sbar}{\bar{S}}

\newcommand{\Qzpre}{Q^{\mathrm{z}}_{\mathrm{pre}}} %
\newcommand{\hV}{V^{\mathrm{pre}}}                 %
\newcommand{\KL}{K_{L}^{\mathrm{pre}}}             %
\newcommand{\KR}{K_{R}}                            %
\newcommand{\KLex}{K_{L}}                          %
\newcommand{\Vex}{V}                               %
\newcommand{\Sfull}{S}                             %
\newcommand{\Gtab}{G}                              %
\newcommand{\hG}{\hat{G}}                          %
\newcommand{\hLam}{\hat{\Lambda}}                  %
\newcommand{\PKL}{\mathrm{PK}_{L}}                 %
\newcommand{\PKR}{\mathrm{PK}_{R}}                 %

\newcommand{\spr}{\widehat{Z}^{\mathrm{out}}}                    %
\newcommand{\Cpre}{C_{\mathrm{pre}}}               %

\DeclareRobustCommand{\Qring}{Q}%
\newcommand{\Ql}{Q^{\mathrm{l}}}                 %
\newcommand{\Qlx}{Q^{\mathrm{lx}}}               %
\newcommand{\Qgrseed}{\Qring^{\mathrm{gr}}}      %
\newcommand{\Qgrz}{Q^{\mathrm{grz}}}             %
\newcommand{\Qpgseed}{\Qring^{\mathrm{pg}}}      %
\newcommand{\Qpg}{Q^{\mathrm{pg}}}               %
\newcommand{\Ast}{\mathscr{A}}

\newcommand{\Bst}{\Qpg}

\DeclareRobustCommand{\Qtri}{\accentset{\scriptscriptstyle\triangle}{Q}}  %
\newcommand{\Qtgls}{\Qtri^{\mathrm{gls}}}
\newcommand{\Qtgrs}{\Qtri^{\mathrm{grs}}}

\newcommand{\GT}{\mathrm{GT}}   %

\providecommand{\Qx}{Q}%
\newif\ifpicto \pictotrue
\ifpicto
\providecolor{stemA}{HTML}{3B4FA0}\providecolor{stemB}{HTML}{B0412E}\providecolor{dpmline}{HTML}{4E8A43}
\usepgfmodule{nonlineartransformations}%
\input{vq_glyphs}
\makeatletter
\newlength{\VQ@w}
\def\VQsp{6.5}%
\def\VQpad{1.6}\def\VQys{0.8}\def\VQrise{3.9}\def\VQlaby{-2.7}\def\VQlabx{7.5}\def\VQlab{\scriptstyle}%
\newcommand{\VQ@draw}[3]{%
  \def\VQ@labs{#3}%
  \ifx\VQ@labs\@empty \def\VQ@S{5.2}\else
    \xdef\VQ@S{\VQsp}%
    \foreach \VQ@l in {#3}{\settowidth{\VQ@w}{$\VQlab\VQ@l$}%
      \pgfmathsetmacro\VQ@new{max(\VQ@S,\VQ@w/1pt+\VQpad)}\xdef\VQ@S{\VQ@new}}%
  \fi
  \mathord{\tikz[baseline=0pt,x=1bp,y=1bp,line cap=butt]{%
    \node[anchor=base west,inner sep=0pt] at (0,0) {$#2$};%
    \begin{scope}[shift={(8.8,\VQrise)},xscale=\VQ@S,yscale=-\VQys*\csname VQy@#1\endcsname]\csname VQg@#1\endcsname\end{scope}%
    \ifx\VQ@labs\@empty\else
      \foreach \VQ@l [count=\VQ@k from 0] in {#3}{\node[anchor=base west,inner sep=0pt] at ({\VQlabx+\VQ@S*\VQ@k},\VQlaby) {$\VQlab\VQ@l$};}%
    \fi
    \path ({8.8+\VQ@S*(\csname VQn@#1\endcsname-1)+(\ifx\VQ@labs\@empty 1.2\else 4.2\fi)},0);}}}
\let\VQ@drawU\VQ@draw
\newcommand{\VQ@map}{%
  \pgfmathsetmacro\VQ@u{\pgf@x/1pt}%
  \pgfmathtruncatemacro\VQ@j{min(max(floor(\VQ@u),0),\VQ@last-1)}%
  \pgfmathtruncatemacro\VQ@jj{\VQ@j+1}%
  \edef\VQ@ba{\csname VQ@B\VQ@j\endcsname}\edef\VQ@bb{\csname VQ@B\VQ@jj\endcsname}%
  \pgfmathsetmacro\VQ@X{\VQ@ba+(\VQ@u-\VQ@j)*(\VQ@bb-\VQ@ba)}%
  \pgf@x=\VQ@X bp\relax}
\newcommand{\VQ@drawC}[3]{%
  \def\VQ@labs{#3}%
  \ifx\VQ@labs\@empty\VQ@drawU{#1}{#2}{}\else
  \settowidth{\VQ@w}{$\VQlab,$}\pgfmathsetmacro\VQ@g{\VQ@w/1bp}%
  \xdef\VQ@L{\VQlabx}%
  \foreach \VQ@l [count=\VQ@k from 0] in {#3}{\settowidth{\VQ@w}{$\VQlab\VQ@l$}\pgfmathsetmacro\VQ@ww{\VQ@w/1bp}%
    \pgfmathsetmacro\VQ@b{\VQ@L+\VQ@ww/2}\expandafter\xdef\csname VQ@B\VQ@k\endcsname{\VQ@b}%
    \expandafter\xdef\csname VQ@Lp\VQ@k\endcsname{\VQ@L}%
    \pgfmathsetmacro\VQ@e{\VQ@L+\VQ@ww}\xdef\VQ@end{\VQ@e}%
    \pgfmathsetmacro\VQ@nL{\VQ@L+\VQ@ww+\VQ@g}\xdef\VQ@L{\VQ@nL}\xdef\VQ@last{\VQ@k}}%
  \mathord{\tikz[baseline=0pt,x=1bp,y=1bp,line cap=butt]{%
    \node[anchor=base west,inner sep=0pt] at (0,0) {$#2$};%
    \begin{scope}\pgftransformnonlinear{\VQ@map}%
      \begin{scope}[x=1pt,shift={(0,\VQrise)},yscale=-\VQys*\csname VQy@#1\endcsname]\csname VQg@#1\endcsname\end{scope}%
    \end{scope}%
    \foreach \VQ@l [count=\VQ@k from 0] in {#3}{\node[anchor=base west,inner sep=0pt] at (\csname VQ@Lp\VQ@k\endcsname,\VQlaby) {$\VQlab\VQ@l$};}%
    \path (\VQ@end+1.2,0);}}\fi}
\newcommand{\VQcommarule}{\let\VQ@draw\VQ@drawC}
\def\VQys{1.15}\def\VQrise{4.4}\VQcommarule%
\def\VQ@firstx#1,#2\@nil{\def\VQ@first{#1}}
\newcommand{\VQ@AB}[4]{\VQ@firstx#4,\@nil\def\VQ@x{x}\ifx\VQ@first\VQ@x\VQ@draw{#2}{#3}{#4}\else\VQ@draw{#1}{#3}{#4}\fi}
\newcommand{\VQ@decl}[3]{%
  \expandafter\RenewDocumentCommand\csname#1\endcsname{e{_}}{\IfValueTF{##1}{\VQ@draw{#2}{#3}{##1}}{\VQ@draw{#2}{#3}{}}}}
\VQ@decl{Qx}{q}{Q}\VQ@decl{Qb}{b}{Q}\VQ@decl{Qm}{m}{Q}\VQ@decl{Qz}{z}{Q}\VQ@decl{Qp}{p}{Q}\VQ@decl{Qg}{g}{Q}
\VQ@decl{Qgls}{gls}{Q}\VQ@decl{Qgrs}{grs}{Q}\VQ@decl{Qtgls}{glsA}{Q}\VQ@decl{Qtgrs}{grsA}{Q}
\VQ@decl{Qlx}{lx}{Q}\VQ@decl{Qgrseed}{grseed}{Q}\VQ@decl{Qgrz}{grz}{Q}\VQ@decl{Qpg}{pg}{Q}
\RenewDocumentCommand{\Qgl}{e{_}}{\IfValueTF{#1}{\VQ@AB{glA}{glB}{Q}{#1}}{\VQ@draw{glA}{Q}{}}}
\RenewDocumentCommand{\Qgr}{e{_}}{\IfValueTF{#1}{\VQ@AB{grA}{grB}{Q}{#1}}{\VQ@draw{grA}{Q}{}}}
\NewDocumentCommand{\QgrB}{e{_}}{\IfValueTF{#1}{\VQ@draw{grB}{Q}{#1}}{\VQ@draw{grB}{Q}{}}}%
\NewDocumentCommand{\QglB}{e{_}}{\IfValueTF{#1}{\VQ@draw{glB}{Q}{#1}}{\VQ@draw{glB}{Q}{}}}
\makeatother
\else
\newcommand{\QgrB}{\Qgr}\newcommand{\QglB}{\Qgl}
\fi

\makeatletter
\renewcommand\section{\@startsection{section}{1}{\z@}{-8pt plus -2pt minus -1pt}{4pt plus 1pt}{\normalfont\large\bfseries}}
\renewcommand\subsection{\@startsection{subsection}{2}{\z@}{-6pt plus -2pt minus -1pt}{3pt plus 1pt}{\normalfont\normalsize\bfseries}}
\makeatother

\ifdefined\submission\fi

\newcommand{\papertitle}{\LN: Linear-Time Approximation of the\\Pseudoknot Partition Function}
\newcommand{\coderepo}{\url{https://github.com/gautammilan/LinearKnot}}
\newcommand{\paperkeywords}{RNA secondary structure; pseudoknots; partition function; base-pairing probabilities; linear-time algorithms}

\title{\papertitle}
\author{%
  Milan Gautam\textsuperscript{1}\quad
  David H.~Mathews\textsuperscript{3}\quad
  Liang Huang\textsuperscript{1,2,$\dagger$}\\[4pt]
  \small\textsuperscript{1}School of EECS, Oregon State University, Corvallis, OR 97331, USA\\
  \small\textsuperscript{2}Dept.~of Biochemistry \& Biophysics, Oregon State University, Corvallis, OR 97331, USA\\
  \small\textsuperscript{3}Dept.~of Biochemistry \& Biophysics, University of Rochester Medical Center, Rochester, NY 14642, USA\\
  \small$^\dagger$Corresponding author
}
\date{}

\begin{document}

\maketitle
\thispagestyle{empty}
\begin{abstract}
Pseudoknots are a common RNA structure where two or more helices cross, and they are central to many functional RNAs, including frameshifting signals, telomerase RNA, riboswitches and viral genomes. However, predicting RNA structures with pseudoknots is notoriously hard,
so the vast majority of RNA secondary structure prediction algorithms and tools only considered pseudoknot-free structures.
With arbitrary pseudoknots, finding the minimum free energy (MFE) structure is NP-hard, and computing the partition function is even harder. Dirks and Pierce (\DPthree) pioneered the first pseudoknot partition function, which underlies \NUPACK's pseudoknot model, but it takes $\order(n^5)$ time and $\order(n^4)$ memory, making it only practical for very short sequences. While the much simpler pseudoknot-free prediction can run in linear time with approximate dynamic programming like \LinearFold and \LinearPartition, these techniques do not transfer easily to pseudoknotted prediction like \DPthree: a partial pseudoknot leaves a gap that is filled by its crossing partner, whose complementary sequence may still lie ahead of the scan, so it cannot be ranked by its inside weight. Even with an estimate of this missing weight, directly pruning \DPthree remains expensive: scoring each partial stem sums over the unspecified boundaries of its partner's complementary sequence. We present \LN, a linear-time approximation of \DPthree in three passes: (a) a \LinearPartition{} calculation gives pseudoknot-free weights; (b) a second pass builds tables that estimate each partial stem's crossing partner; and (c) a third scans a left-to-right reformulation of \DPthree that reserves the 5$'$ nucleotides in the pseudoknot helix for subsequent pairing, keeping the states with the largest estimated contribution. Without pruning, this reformulation computes exactly \DPthree's pseudoknot recursion, over the same structures with the same energies. Kept states are combined by \DPthree's exact recursions, so the only approximation is which states the beam keeps. The scan of \LN is linear in time and memory; end-to-end, time grows linearly ($n^{1.01}$) and memory marginally above linear ($n^{1.15}$) from 1{,}000 to 9{,}392\nt, because the supporting passes are not yet fully beam-bounded. It folds the 9{,}392\nt Rous sarcoma virus genome in 2.4 hours and 90.5\,GB, where exact \DPthree would need an estimated 40{,}000 years and 54 million GB of memory (extrapolating \NUPACK's measured $n^{5.39}$ time and $n^{3.91}$ memory), and matches exact \DPthree base-pair probabilities on 418 \PseudoBase RNAs with a median RMSD of $9.2\times10^{-7}$ (mean $6.8\times10^{-5}$).
Our code is available at \coderepo.

\end{abstract}
\newpage
\setcounter{page}{1}

\section{Introduction}\label{sec:intro}

RNAs play central roles in gene regulation, catalysis and viral replication, and their functions are tied to their structures. Determining structures experimentally is slow and costly, so computational prediction of secondary structure is widely used. Over the past few decades, secondary structure prediction has improved significantly in terms of accuracy and efficiency~\cite{reuter2010rnastructure,huang2019linearfold}, where the standard method is dynamic programming with nearest-neighbor energy parameters.

However, one important type of structures called ``pseudoknots'' remains extremely challenging, where base pairs cross each other, e.g., $(i,j)$ and $(k,l)$ where $i\!<\!k\!<\!j\!<\!l$. Pseudoknots are central to ribosomal frameshifting, telomerase RNA, riboswitches and the replication of many viral genomes~\cite{staple2005,brierley2007}. They are much harder to predict than pseudoknot-free structures because these crossing pairs are beyond the expressive capacity of standard cubic-time dynamic programming; in fact, finding the MFE structure with arbitrary pseudoknots is NP-hard under energy-based models~\cite{lyngso2000}.

Existing algorithms for pseudoknot prediction fall into two broad categories. The first camp is {\bf supercubic-time dynamic programming}. They aim to solve particular subclasses of pseudoknots with exact dynamic programming, at the cost of very high time complexity, ranging from $\order(n^4)$ to $\order(n^6)$, with increasingly larger subclasses of predictable structures. Examples include Rivas and Eddy~\cite{rivas1999} (most expressive, $\order(n^6)$ time), Akutsu~\cite{akutsu2000}, Uemura et al.~\cite{uemura1999} and pknotsRG~\cite{reeder2004} (simpler pseudoknots, $\order(n^4)$ to $\order(n^5)$ time), and Dirks and Pierce (\DPthree)~\cite{dirks2003} in $\order(n^5)$ time. Knotty~\cite{jabbari2018} finds the MFE structure exactly in the more general CCJ class in $\order(n^5)$ time; sparsification reduces its space from CCJ's $\Theta(n^4)$ to $\Theta(n^3+Z)$, where $Z$ is the number of candidates. These methods model pseudoknots directly, which is principled, but is prohibitively slow and not practical at all.

On the other hand, the second camp uses {\bf heuristic methods} based on pseudoknot-free predictions from $\order(n^3)$-time dynamic programming. Examples include \HotKnots~\cite{ren2005} that searches for low-energy structures, ProbKnot~\cite{bellaousov2010}, \ThreshKnot~\cite{zhang2019threshknot} and \IPknot~\cite{sato2011} that assemble pseudoknots from pseudoknot-free pair probabilities, and HFold~\cite{jabbari2007} and CParty~\cite{gray2025cparty} that fold pseudoknots around a given pseudoknot-free structure. These heuristic algorithms run much faster (most build on $\order(n^3)$ pseudoknot-free folding; \IPknot also solves an integer program), but with very limited accuracy because (a) they do not model pseudoknots directly and (b) their base predictions are computed under pseudoknot-free assumptions.

Therefore, both camps suffer from severe limitations. To make things worse, an RNA folds into an equilibrium ensemble of alternative structures. The partition function over this ensemble yields base-pairing probabilities~\cite{mccaskill1990}, which are more informative than a single minimum free energy structure~\cite{zuker1981,mathews1999} and feed downstream methods such as maximum expected accuracy decoding~\cite{do2006contrafold}, \ThreshKnot~\cite{zhang2019threshknot} and stochastic sampling~\cite{ding2003sampling}. With pseudoknots, the ensemble must include crossing structures. \DPthree{} provided the first partition function algorithm with pseudoknots. Its class covers H-type pseudoknots nested inside multiloops and inside other pseudoknots, and it is available in \NUPACK~3 as its pseudoknot option for single-stranded RNA~\cite{zadeh2011,nupack3guide}.

However, \DPthree costs $\order(n^5)$ time and $\order(n^4)$ memory. In our runs \NUPACK takes 18.8 hours and 283\,GB of memory for a 418\nt RNA, and its measured scaling ($n^{5.39}$ time, $n^{3.91}$ memory) gives about 2 years and 42\,TB for a 1{,}500\nt 16S rRNA. Most structured RNAs that contain pseudoknots, such as RNase P RNAs, ribosomal RNAs and viral genomes, are therefore out of reach, and the predictions of the \DPthree model on long RNAs have been hard to examine.

Approximating a pseudoknot partition function is harder than the pseudoknot-free case that \LinearPartition addresses. Because the number of possible base pairs grows with the square of the sequence length, a linear-time approximation requires pruning the space of considered pairs; this pruning is more difficult for structures with pseudoknots. A pseudoknot-free state covers its whole prefix, so states at the same position differ only in what they have built; a partial pseudoknot leaves a gap that only a partner read later can fill, so it cannot be ranked for possible pruning by its prefix weight alone. Moreover, even with an estimate of this missing weight, scoring a \DPthree partial stem is expensive: it sums over the unrecorded boundaries of its partner's complementary sequence and over both stem roles (\cref{sec:beamdp}). To our knowledge, no linear-time approximation of a pseudoknot partition function exists; the closest prior work, CParty~\cite{gray2025cparty}, computes one in cubic time, conditional on a given pseudoknot-free structure.

We address both problems with an algorithm called \LN, a linear-time approximation of the \DPthree partition function and base-pairing probabilities. Like \LinearFold and \LinearPartition~\cite{huang2019linearfold,zhang2020linearpartition}, it scans from 5$'$ to 3$'$ with beam pruning and is global, with no limit on the span of a base pair or a pseudoknot. To compare partial pseudoknots fairly, we estimate the weight of the regions not yet built, the outside weight, from pseudoknot-free tables computed before the scan (\cref{sec:overview}). To remove both sums, we reserve the partner's complementary sequence before building it. Each state then records where this sequence lies and builds only one stem, whose role is fixed, so its rank score no longer sums over placements or roles (\cref{sec:pass3}). The partition function still includes every placement and role, but now each one is a separate state, which the beam can score and prune on its own. \DPthree's recursions form a grammar, whose rules combine smaller states into larger ones. Without pruning, our transformed grammar computes exactly \DPthree's partition function; the only approximation is which states the beam keeps.

\LN extends \LinearFold~\cite{huang2019linearfold} and \LinearPartition~\cite{zhang2020linearpartition}, which bring left-to-right beam search from incremental parsing~\cite{huang2010} to pseudoknot-free folding and rank states by their prefix weight. However, a partial pseudoknot leaves a gap that its prefix weight cannot score, so we rank by an estimate of the outside weight, in the spirit of A*~\cite{klein2003} and coarse-to-fine parsing~\cite{charniak2006}, where a simpler grammar supplies outside estimates that prune the states of a richer one; here the pseudoknot-free model of Passes~1 and~2 provides the simpler grammar. Our reformulation of \DPthree uses two exact program transformations: \emph{folding}, of which \DPthree's own $\order(n^5)$ evaluation and the hook trick~\cite{eisner1999} are instances, and \emph{speculation}~\cite{eisner2007}, which lets the scan start a pseudoknot before its partner is read.

\section{Background: Pseudoknots and the DP03 Recursions}\label{sec:prelim}

\paragraph{Energy model.} For a sequence $\vecx=x_1\cdots x_n$, the partition function is $\Zfn=\sum_{\vecy\in\Yset}e^{-\Gfe(\vecy)/\RT}$ over the structures $\Yset$ in \DPthree's class, with free energy $\Gfe(\vecy)$, gas constant $R$ and temperature $T$. Nested loops use the nearest-neighbor model~\cite{mathews1999}, and a multiloop costs $\alphaone+\alphatwo B+\alphathree U$ for $B$ pairs bordering it and $U$ unpaired bases inside it. \DPthree scores a pseudoknot with the same type of linear approximation~\citep[Eq.~16]{dirks2003}, $\betaone+\betatwo B^{p}+\betathree U^{p}$, with $B^{p}$ the pairs that border the pseudoknot interior and $U^{p}$ the unpaired bases inside it; $\betaone$ becomes $\betaonem$ inside a multiloop and $\betaonep$ inside another pseudoknot. We write $\omegaone=e^{-\betaone/\RT}$, $\omegaonem=e^{-\betaonem/\RT}$ and $\omegaonep=e^{-\betaonep/\RT}$.

\begin{figure}[t]
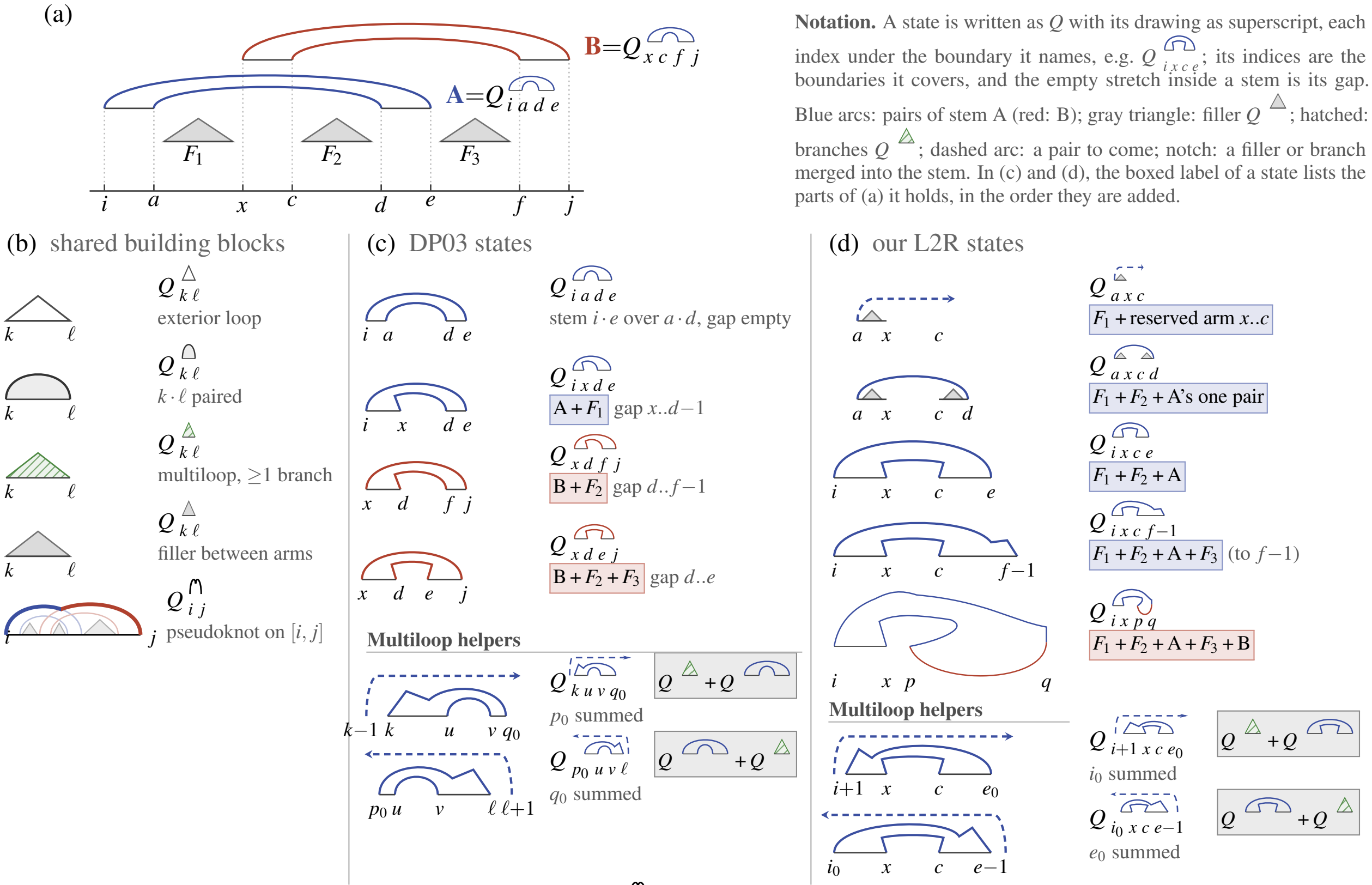

\centering
\begin{minipage}[c]{0.54\linewidth}\centering\input{figs/dp03fig_defs}%
\input{figs/l2rfig_defs}%
\begin{tikzpicture}[x=1bp, y=-1bp]
\node[anchor=base west, inner sep=0pt, font=\small] at (-14,8) {(a)};
\path[draw=black!25, line width=0.5bp, dash pattern=on 0.5bp off 1bp] (7.03,66.14) -- (7.03,37.70);
\path[draw=black!25, line width=0.5bp, dash pattern=on 0.5bp off 1bp] (24.04,66.14) -- (24.04,37.70);
\path[draw=black!25, line width=0.5bp, dash pattern=on 0.5bp off 1bp] (54.65,66.14) -- (54.65,21.03);
\path[draw=black!25, line width=0.5bp, dash pattern=on 0.5bp off 1bp] (71.66,66.14) -- (71.66,21.03);
\path[draw=black!25, line width=0.5bp, dash pattern=on 0.5bp off 1bp] (102.28,66.14) -- (102.28,37.70);
\path[draw=black!25, line width=0.5bp, dash pattern=on 0.5bp off 1bp] (119.29,66.14) -- (119.29,37.70);
\path[draw=black!25, line width=0.5bp, dash pattern=on 0.5bp off 1bp] (149.90,66.14) -- (149.90,21.03);
\path[draw=black!25, line width=0.5bp, dash pattern=on 0.5bp off 1bp] (166.91,66.14) -- (166.91,21.03);
\path[draw=black!75, line width=0.5bp] (1.93,66.14) -- (172.01,66.14);
\path[draw=black!75, line width=0.5bp] (7.03,66.14) -- (7.03,63.87);
\path[draw=black!75, line width=0.5bp] (24.04,66.14) -- (24.04,63.87);
\path[draw=black!75, line width=0.5bp] (54.65,66.14) -- (54.65,63.87);
\path[draw=black!75, line width=0.5bp] (71.66,66.14) -- (71.66,63.87);
\path[draw=black!75, line width=0.5bp] (102.28,66.14) -- (102.28,63.87);
\path[draw=black!75, line width=0.5bp] (119.29,66.14) -- (119.29,63.87);
\path[draw=black!75, line width=0.5bp] (149.90,66.14) -- (149.90,63.87);
\path[draw=black!75, line width=0.5bp] (166.91,66.14) -- (166.91,63.87);
\path[fill=black!14, draw=black!55, line width=0.5bp] (27.44,49.13) -- (39.35,41.19) -- (51.25,49.13) -- cycle;
\path[draw=black!75, line width=0.5bp] (27.44,49.13) -- (51.25,49.13);
\path[fill=white] (24.77,59.43) -- (53.92,59.43) -- (53.92,50.16) -- (24.77,50.16) -- cycle;
\path[fill=black!14, draw=black!55, line width=0.5bp] (75.06,49.13) -- (86.97,41.19) -- (98.88,49.13) -- cycle;
\path[draw=black!75, line width=0.5bp] (75.06,49.13) -- (98.88,49.13);
\path[fill=white] (72.41,59.43) -- (101.53,59.43) -- (101.53,50.16) -- (72.41,50.16) -- cycle;
\path[fill=black!14, draw=black!55, line width=0.5bp] (122.69,49.13) -- (134.59,41.19) -- (146.50,49.13) -- cycle;
\path[draw=black!75, line width=0.5bp] (122.69,49.13) -- (146.50,49.13);
\path[fill=white] (119.96,59.43) -- (149.22,59.43) -- (149.22,50.17) -- (119.96,50.17) -- cycle;
\path[draw=stemA, line width=0.75bp] (7.03,37.70) .. controls (7.03,22.59) and (119.29,22.59) .. (119.29,37.70);
\path[draw=stemA, line width=0.75bp] (24.04,37.70) .. controls (24.04,27.87) and (102.28,27.87) .. (102.28,37.70);
\path[draw=black!75, line width=0.5bp] (7.03,37.70) -- (24.04,37.70);
\path[draw=black!75, line width=0.5bp] (102.28,37.70) -- (119.29,37.70);
\path[draw=stemB, line width=0.75bp] (71.66,21.03) .. controls (71.66,11.19) and (149.90,11.19) .. (149.90,21.03);
\path[draw=stemB, line width=0.75bp] (54.65,21.03) .. controls (54.65,4.50) and (166.91,4.50) .. (166.91,21.03);
\path[draw=black!75, line width=0.5bp] (54.65,21.03) -- (71.66,21.03);
\path[draw=black!75, line width=0.5bp] (149.90,21.03) -- (166.91,21.03);
\node[anchor=base west, inner sep=0pt, font=\footnotesize] at (5.72,72.99) {$i$};
\node[anchor=base west, inner sep=0pt, font=\footnotesize] at (21.84,71.28) {$a$};
\node[anchor=base west, inner sep=0pt, font=\footnotesize] at (52.46,73.30) {$x$};
\node[anchor=base west, inner sep=0pt, font=\footnotesize] at (69.69,71.28) {$c$};
\node[anchor=base west, inner sep=0pt, font=\footnotesize] at (99.88,73.30) {$d$};
\node[anchor=base west, inner sep=0pt, font=\footnotesize] at (117.31,71.28) {$e$};
\node[anchor=base west, inner sep=0pt, font=\footnotesize] at (147.41,73.17) {$f$};
\node[anchor=base west, inner sep=0pt, font=\footnotesize] at (164.90,72.99) {$j$};
\node[anchor=base west, inner sep=0pt, font=\footnotesize] at (33.9,55.76) {$F_1$};
\node[anchor=base west, inner sep=0pt, font=\footnotesize] at (81.5,55.76) {$F_2$};
\node[anchor=base west, inner sep=0pt, font=\footnotesize] at (129.1,55.76) {$F_3$};
\node[anchor=base west, inner sep=0pt, font=\small] at (124.38,35.92) {\textcolor{stemA}{\textbf{A}}${=}\Qg_{i,a,d,e}$};
\node[anchor=base west, inner sep=0pt, font=\small] at (172.01,19.19) {\textcolor{stemB}{\textbf{B}}${=}\Qg_{x,c,f,j}$};
\end{tikzpicture}%
\end{minipage}\hfill
\begin{minipage}[c]{0.42\linewidth}\scriptsize\color{black!65}\textbf{Notation.} A state is written as $Q$ with its drawing as superscript, each index under the boundary it names, e.g.\ $\Qgr_{i,x,c,e}$; its indices are the boundaries it covers, and the empty stretch inside a stem is its gap. Blue arcs: pairs of stem A (red: B); gray triangle: filler $\Qz$; hatched: branches $\Qm$; dashed arc: a pair to come; notch: a filler or branch merged into the stem. In (c) and (d), the boxed label of a state lists the parts of (a) it holds, in the order they are added.\end{minipage}\par\vspace{5pt}
\input{figs/fig_states}\vspace{-6pt}%
\caption{(a)~The five factors of \DPthree's pseudoknot rule $\Qp$ (\cref{eq:Qp}) for an H-type pseudoknot: stem A (blue) has outer pair $\pair{i}{e}$ and inner pair $\pair{a}{d}$, stem B (red) has $\pair{x}{j}$ and $\pair{c}{f}$, both gap states $\Qg$; the fillers $\Fone=\Qz_{a+1,x-1}$, $\Ftwo=\Qz_{c+1,d-1}$, $\Fthree=\Qz_{e+1,f-1}$ are full partition functions, so they may contain further pseudoknots. (b)~Building blocks shared by both algorithms, (c)~\DPthree's states, and (d)~our L2R states (\cref{sec:pass3}), in the order the scan creates them, with indices aligned by boundary. In (d), B's reserved arm $x..c$ is an empty gap that B's arm will fill; A's seed $\Qgrseed_{a,x,c,d}$, a stem of one pair, grows into $\Qgr_{i,x,c,e}$ (Step~3 of \cref{fig:l2rsteps}). The key difference is in the stem with both fillers: \DPthree builds it on stem B ($\QgrB_{x,d,e,j}$), so its gap is A's right arm $d..e$; we build it on stem A ($\Qgr_{i,x,c,e}$), so its gap is B's left arm $x..c$; this way A records where B's arm lies and has a single role, which removes both sums from its rank score (\cref{sec:beamdp}). In $\Qgls$ and $\Qgrs$, the dashed arc is the multiloop closing pair the state will receive, and the branches extend the stem to its left or right. Every state in (d) ends at the scan position.}
\label{fig:htype}\label{fig:states}\label{fig:dp03states}\label{fig:l2rstates}
\end{figure}

\begin{figure}[!b]
\centering
\resizebox{0.95\linewidth}{!}{\input{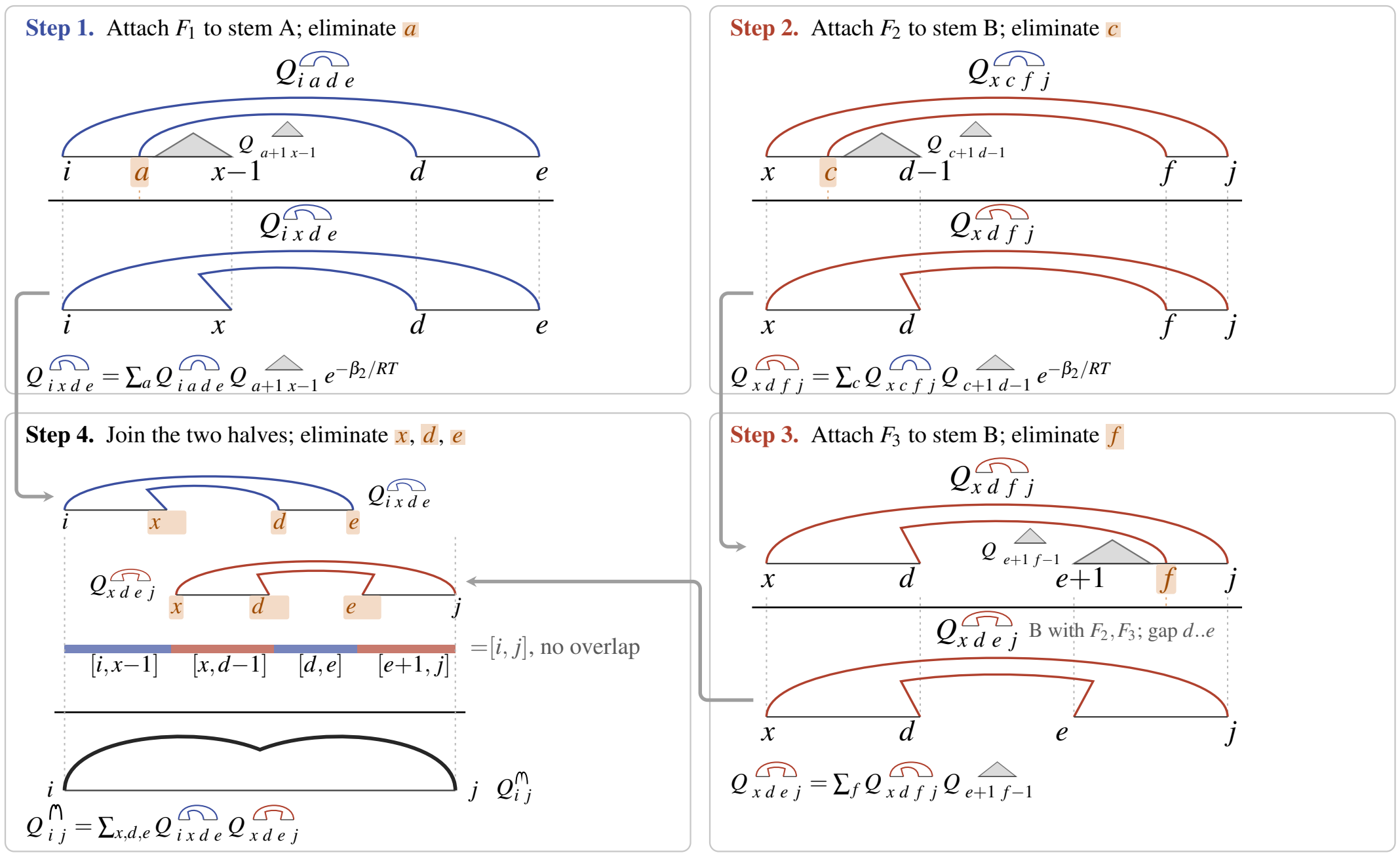}}
\caption{How \DPthree evaluates the pseudoknot rule $\Qp$ (\cref{eq:Qp}) in $\order(n^5)$ (pieces as in \cref{fig:htype}a). Each step multiplies the items above its bar and sums over the highlighted boundary indices, storing the item below; gray arrows carry a stored item to the step that uses it. Steps~1--3 build the two halves, Step~4 joins them into $\Qp$ (outline); $e^{-\betatwo/\RT}$ is \DPthree's penalty per inner pair.}
\label{fig:pkassembly}
\end{figure}
\paragraph{Tables.} \DPthree computes $\Zfn$ with six tables (\cref{fig:dp03states}b,c): $\Qx_{k,\ell}$, the partition function of the interval $[k,\ell]$ in the exterior loop; $\Qb_{k,\ell}$, the same with $\pair{k}{\ell}$ paired; $\Qm_{k,\ell}$, a region inside a multiloop with at least one branch; $\Qz_{k,\ell}$, a \emph{filler}, a region between two stem arms inside a pseudoknot ($\Fone,\Ftwo,\Fthree$ in \cref{fig:htype}a); $\Qp_{i,j}$, a pseudoknot that spans exactly $[i,j]$; and the \emph{gap state} $\Qg_{p,u,v,q}$, a stem with outer pair $\pair{p}{q}$ and inner pair $\pair{u}{v}$ whose gap $[u{+}1,v{-}1]$ is left empty. $\Qx$, $\Qb$ and $\Qm$ are the pseudoknot-free tables, extended with pseudoknot cases; $\Qz$ plays the role of $\Qx$ between two stem arms, but with \DPthree's pseudoknot penalties, $\betatwo$ per bordering pair and $\betathree$ per unpaired base, in place of the multiloop's $\alphatwo$ and $\alphathree$. The exterior and filler tables each gain one case for a rightmost pseudoknot $\Qp_{i,j}$:
\begin{align}
\Qx_{k,\ell} &= 1+\sum_{k\le i<j\le \ell} \Qx_{k,i-1}\,\big(\Qb_{i,j}+\omegaone\,\Qp_{i,j}\big), \label{eq:Q}\\
\Qz_{k,\ell} &= e^{-\betathree(\ell-k+1)/\RT}+\sum_{k\le i<j\le \ell}\Qz_{k,i-1}\,\big(e^{-\betatwo/\RT}\,\Qb_{i,j}+\omegaonep\,e^{-2\betatwo/\RT}\,\Qp_{i,j}\big)\,e^{-\betathree(\ell-j)/\RT}, \label{eq:Qz}
\end{align}
where $j$ is the last paired base of the interval, the bases $j{+}1..\ell$ are unpaired, and $\Qx_{k,k-1}=\Qz_{k,k-1}=1$. Inside a pseudoknot, each pair that borders its interior costs $\betatwo$ (energy model above). A nested pair $\Qb_{i,j}$ in a filler has one such pair, $\pair{i}{j}$, hence the factor $e^{-\betatwo/\RT}$ in the filler recursion $\Qz$ (\cref{eq:Qz}); a pseudoknot $\Qp_{i,j}$ in a filler has two, the outer pairs of its two stems, hence $e^{-2\betatwo/\RT}$. A finished pseudoknot enters $\Qb$ and $\Qm$ as a multiloop branch, with $\omegaonem$.

A gap state grows outward one loop at a time. Let $\Nin(p,q)$ be the set of pairs $\pair{p_0}{q_0}$ with $p<p_0<q_0<q$ such that $\pair{p}{q}$ closes a loop around $\pair{p_0}{q_0}$ inside the spanning region: a stack or interior loop with at most $\Lmax=30$ unpaired bases, or a multiloop whose other branches lie in $\Qm$ regions to the left and/or right of $\pair{p_0}{q_0}$. Let $\wloop{p}{q}{p_0}{q_0}$ be the Boltzmann weight of that loop. Then
\begin{equation}
\Qg_{p,u,v,q} = \ind{p=u}\,\ind{q=v} + \sum_{(p_0,q_0)\in\Nin(p,q)} \wloop{p}{q}{p_0}{q_0}\,\Qg_{p_0,u,v,q_0}, \label{eq:Qg}
\end{equation}
for pairs that can form, where the first term is a single-pair stem. \DPthree evaluates the multiloop part of this sum through two helper tables, $\Qgls$ and $\Qgrs$, which does not change its value (\cref{fig:mlsplit}).

\paragraph{Pseudoknots.} With the indices of \cref{fig:htype}a, $i\le a<x\le c<d\le e<f\le j$, \DPthree defines
\begin{equation}
\Qp_{i,j}=e^{-2\betatwo/\RT}\sum_{a,x,c,d,e,f}\Qg_{i,a,d,e}\,\Qg_{x,c,f,j}\;\Qz_{a+1,x-1}\,\Qz_{c+1,d-1}\,\Qz_{e+1,f-1}, \label{eq:Qp}
\end{equation}
where the factor $e^{-2\betatwo/\RT}$ is \DPthree's penalty $\betatwo$ for each of the two inner pairs $\pair{a}{d}$ and $\pair{c}{f}$, which border the pseudoknot interior. As in \DPthree's recursions~\citep[Figs.~15 and~18]{dirks2003}, $\Qp$ charges only these two; the outer pairs $\pair{i}{e}$ and $\pair{x}{j}$ are charged by the loop that contains the pseudoknot: $e^{-2\betatwo/\RT}$ inside a filler (\cref{eq:Qz}; \citep[Fig.~12]{dirks2003}) and nothing in the exterior loop (\cref{eq:Q}). The initiation penalty $\betaone$ is charged the same way, by the containing loop ($\omegaone$, $\omegaonem$, $\omegaonep$), and not by $\Qp$.  Written as one sum, \cref{eq:Qp} has eight free indices and would cost $\order(n^8)$ time. \DPthree folds it into four steps (\cref{fig:pkassembly}), summing out each index as soon as every factor that uses it has been multiplied in, the principle of the hook trick~\cite{eisner1999}: Steps~1 and~2 attach $\Fone$ to A and $\Ftwo$ to B, Step~3 attaches $\Fthree$ to B, and Step~4 joins the two halves. Each step has at most five free indices, hence $\order(n^5)$ time, and the four-index gap states need $\order(n^4)$ memory. We reindex \DPthree's auxiliary tables so that the index after an attached filler is the first base of the open gap: $\Qgl_{i,x,d,e}$ leaves $[x,d{-}1]$ open and $\Qgr_{x,d,e,j}$ leaves A's arm $[d,e]$, so an arm of the other stem appears directly as indices. A finished pseudoknot enters the exterior loop, multiloops and fillers with $\omegaone$, $\omegaonem$ and $\omegaonep$.

\section{\texorpdfstring{Ranking Partial Pseudoknots}{Ranking Partial Pseudoknots}}\label{sec:ranking}

\begin{figure}[H]%
\begin{minipage}[c]{0.64\linewidth}\centering\resizebox{\linewidth}{!}{\providecolor{stemA}{HTML}{3B4FA0}
\providecolor{passtwo}{HTML}{C9852B}
\providecolor{stemB}{HTML}{B5402E}
\begin{tikzpicture}[x=1cm, y=1cm, font=\footnotesize]
\tikzset{
  bb/.style={black!75, line width=0.5pt},                         %
  st/.style={stemA, line width=1.0pt},                            %
  cp/.style={passtwo, line width=0.9pt, dash pattern=on 3pt off 2pt}, %
  ix/.style={below, inner sep=1.2pt},                             %
  sc/.style={black!40, densely dashed, line width=0.5pt},         %
  strip/.style={line width=2.4pt},                                %
}
\def\arc#1#2#3#4{\draw[#4] (#1,0) .. controls (#1,{4*#3/3}) and (#2,{4*#3/3}) .. (#2,0);}

\node[anchor=west, font=\small] at (-0.1,2.15) {\textbf{(a)} prefix scoring leaves a gap};
\node[anchor=west, font=\scriptsize] at (-0.1,1.7) {prefix score: $\Cpref(i{-}1)\,\Qg_{i,a,d,e}$};
\def\q{4.9}\def\n{7.4}
\draw[sc] (\q,-2.95) -- (\q,1.35) node[above, inner sep=1pt, text=black!55] {scan at $e$};

\begin{scope}[shift={(0,0.15)}]
  \def\p{1.5}\def\u{2.35}\def\v{3.85}\def\xb{2.8}\def\cb{3.35}%
  \node[anchor=west, text=black!65] at (-0.1,1.0) {small gap, $s_1$};
  \draw[bb] (0,0) -- (\n,0);
  \arc{0.3}{0.85}{0.28}{cp}\arc{6.3}{6.95}{0.28}{cp}
  \arc{\xb}{5.9}{0.95}{cp}\arc{\cb}{5.6}{0.72}{cp}%
  \arc{\p}{\q}{0.78}{st}\arc{\u}{\v}{0.45}{st}
  \draw[stemB, line width=2.2pt] (\xb,0) -- (\cb,0);%
  \draw[strip, stemA!55] (0,-0.32) -- (\u,-0.32);  \draw[strip, stemA!55] (\v,-0.32) -- (\q,-0.32);
  \draw[strip, black!22] (\u,-0.32) -- (\v,-0.32); \draw[strip, black!22] (\q,-0.32) -- (\n,-0.32);
  \foreach \x/\l in {0/1,\p/i_1,\u/a_1,\v/d_1,\q/e,\n/n} \node[ix] at (\x,-0.36) {$\l$};
  \node[ix, text=stemB] at (\xb,-0.36) {$x$};\node[ix, text=stemB] at (\cb,-0.36) {$c$};
\end{scope}
\begin{scope}[shift={(0,-2.15)}]
  \def\p{0.95}\def\u{1.5}\def\v{4.35}
  \node[anchor=west, text=black!65] at (-0.1,1.0) {wide gap, $s_2$};
  \draw[bb] (0,0) -- (\n,0);
  \arc{0.15}{0.6}{0.25}{cp}\arc{6.4}{6.95}{0.28}{cp}
  \arc{2.2}{6.05}{1.0}{cp}\arc{2.6}{5.65}{0.8}{cp}
  \arc{\p}{\q}{0.82}{st}\arc{\u}{\v}{0.6}{st}
  \draw[strip, stemA!55] (0,-0.32) -- (\u,-0.32);  \draw[strip, stemA!55] (\v,-0.32) -- (\q,-0.32);
  \draw[strip, black!22] (\u,-0.32) -- (\v,-0.32); \draw[strip, black!22] (\q,-0.32) -- (\n,-0.32);
  \foreach \x/\l in {0/1,\p/i_2,\u/a_2,\v/d_2,\q/e,\n/n} \node[ix] at (\x,-0.36) {$\l$};
\end{scope}

\draw[black!25] (7.85,-3.3) -- (7.85,2.2);

\begin{scope}[shift={(8.3,0)}]
  \def\n{7.4}\def\q{4.9}
  \def\k{0.135}   %
  \def\off{0.42}  %
  \tikzset{
    outfill/.style={fill=passtwo!16, draw=passtwo!80, line width=0.5pt},
    infill/.style={fill=stemA, draw=stemA, line width=0.5pt},
  }
  \def\tri#1#2#3{\fill[#3] (#1,-\off) -- ({(#1+#2)/2},{-\off-(#2-#1)*\k}) -- (#2,-\off) -- cycle;}
  \node[anchor=west, font=\small] at (-0.1,2.15) {\textbf{(b)} inside $\times$ outside covers $1..n$};
  \draw[sc] (\q,-3.75) -- (\q,1.35) node[above, inner sep=1pt, text=black!55] {scan at $e$};

  \begin{scope}[shift={(0,0.15)}]
    \def\p{1.5}\def\u{2.35}\def\v{3.85}
    \draw[bb] (0,0) -- (\n,0);
    \arc{0.3}{0.85}{0.22}{cp}\arc{6.3}{6.95}{0.22}{cp}
    \arc{2.8}{5.9}{0.72}{cp}\arc{3.35}{5.6}{0.55}{cp}%
    \arc{\p}{\q}{0.6}{st}\arc{\u}{\v}{0.34}{st}
    \foreach \x/\l in {0/1,\p/i_1,\u/a_1,\v/d_1,\q/e,\n/n} \node[ix] at (\x,0) {$\l$};
    \tri{0}{\n}{outfill}\tri{\p}{\q}{infill}\tri{\u}{\v}{outfill}
    \node[text=white, font=\scriptsize\bfseries, inner sep=0pt] at ({(\p+\q)/2+0.3},{-\off-0.24}) {inside};
    \node[text=passtwo!85!black, font=\tiny, anchor=west, inner sep=0pt] at (0.25,{-\off-0.09}) {prefix};
    \node[text=passtwo!85!black, font=\tiny, anchor=west, inner sep=0pt] at (\q+0.06,{-\off-0.09}) {partner, growth, suffix};
  \end{scope}
  \begin{scope}[shift={(0,-2.15)}]
    \def\p{0.95}\def\u{1.5}\def\v{4.35}
    \draw[bb] (0,0) -- (\n,0);
    \arc{0.15}{0.6}{0.2}{cp}\arc{6.4}{6.95}{0.22}{cp}
    \arc{2.2}{6.05}{0.75}{cp}\arc{2.6}{5.65}{0.6}{cp}
    \arc{\p}{\q}{0.62}{st}\arc{\u}{\v}{0.45}{st}
    \foreach \x/\l in {0/1,\p/i_2,\u/a_2,\v/d_2,\q/e,\n/n} \node[ix] at (\x,0) {$\l$};
    \tri{0}{\n}{outfill}\tri{\p}{\q}{infill}\tri{\u}{\v}{outfill}
    \node[text=passtwo!85!black, font=\tiny, inner sep=0pt] at ({(\u+\v)/2},{-\off-0.1}) {gap: partner's arm, fillers};
    \node[text=stemA, font=\scriptsize\bfseries, inner sep=0pt, anchor=north] at ({(\p+\q)/2},{-\off-(\q-\p)*\k-0.02}) {inside};
    \node[text=passtwo!85!black, font=\tiny, anchor=west, inner sep=0pt] at (\q+0.06,{-\off-0.09}) {partner, growth, suffix};
  \end{scope}
\end{scope}

\end{tikzpicture}%
}\end{minipage}\hfill
\begin{minipage}[c]{0.34\linewidth}
\captionof{figure}{Ranking partial pseudoknots. Each state is a partial stem A, $\Qg_{i,a,d,e}$, with outer pair $\pair{i}{e}$ and inner pair $\pair{a}{d}$ (letters of \cref{fig:htype}a). (a)~The prefix score covers the prefix and the stem (blue) but not the gap, and two states at the same scan position $e$ leave different gaps. The red bar in $s_1$'s gap is partner B's left arm $x..c$, one of the placements its outside weight sums over (\cref{sec:beamdp}). (b)~The outside weight $\Zout(s)$ covers everything the state has not built (orange); \LN ranks by the rank score $\Zin(s)\,\spr(s)$ (\cref{eq:score}), with $\spr(s)\approx\Zout(s)$.}
\label{fig:gap}
\end{minipage}
\end{figure}
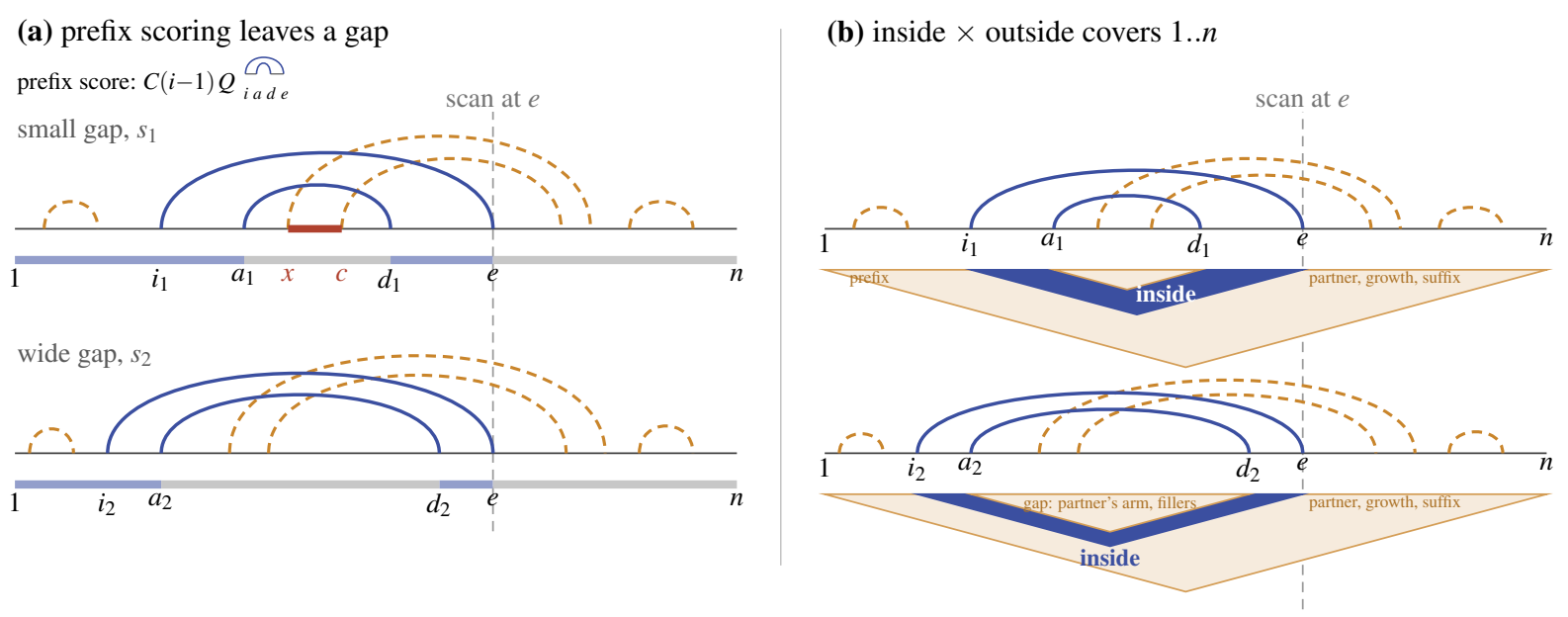\vspace{-8pt}%

\subsection{\texorpdfstring{Ranking partial pseudoknots by their outside weight}{Ranking partial pseudoknots by their outside weight}}\label{sec:target}
\paragraph{The ideal rank score.} A beam search reads the sequence from 5$'$ to 3$'$ and, at each position $j$, keeps only the $b$ states with the largest \emph{rank score}; the other states are discarded. For a state $s$, let $\Zin(s)$ be its \emph{inside weight}, the total weight of the substructures that $s$ covers (blue in \cref{fig:gap}b), and $\Zout(s)$ its \emph{outside weight}, the total weight of all ways to complete $s$ into a full structure of $x_1\cdots x_n$ (orange in \cref{fig:gap}b). Their product $\Zin(s)\,\Zout(s)$ is the total weight of all structures that contain $s$, which is exactly the part of $\Zfn$ that is lost if $s$ is discarded. The ideal rank score for beam pruning is therefore $\Zin(s)\,\Zout(s)$. The scan cannot compute it, because $\Zout(s)$ depends on the unread bases after $j$.

\paragraph{\LinearPartition's rank score.} For a nested state $\Qb_{i,j}$ in the exterior loop, the outside weight is the weight of the prefix before $i$ times the weight of the suffix after $j$: $\Zout(\Qb_{i,j})=\Cpref(i{-}1)\,\Sfull(j{+}1)$, with the \emph{prefix weight} $\Cpref(k)=Q_{1,k}$ and the \emph{suffix weight} $\Sfull(k)=Q_{k,n}$. The scan has read the prefix, so it knows $\Cpref(i{-}1)$. It has not read the suffix, but every state that ends at $j$ is multiplied by the same factor $\Sfull(j{+}1)$, so dropping this factor does not change which states have the largest rank scores. \LinearPartition therefore ranks every state that ends at $j$ by its \emph{prefix score} $\Cpref(i{-}1)\,\Qb_{i,j}$.

\paragraph{Why the prefix score fails for a gap state $\Qg$.} A stem of a partial pseudoknot is a gap state, here stem A, $\Qg_{i,a,d,e}$ (\cref{fig:states}c): it has outer pair $\pair{i}{e}$, inner pair $\pair{a}{d}$, and a gap $[a{+}1,d{-}1]$ that the state leaves empty. In a full structure the gap is not empty. Consider a gap state that will serve as stem A: its gap holds B's left arm and the first and second fillers $\Fone,\Ftwo$, while B's right arm lies after $e$, where the scan has not read yet (\cref{fig:htype}a); a stem that will serve as B holds A's right arm in its gap instead (\cref{tab:openidx}). \Cref{fig:gap}a shows two gap states that end at the same scan position $e$, $s_1=\Qg_{i_1,a_1,d_1,e}$ with a small gap and $s_2=\Qg_{i_2,a_2,d_2,e}$ with a wide gap. Their prefix scores cover $[1,e]$ \emph{except} their gaps, and the gaps differ, so each score misses a different weight. Unlike the nested suffix, this missing weight is not shared, and dropping it changes which state ranks higher: the comparison is no longer fair.

\paragraph{The rank score of \LN.} \LN estimates, for every state $s$, its whole outside weight, gap included: $\spr(s)\approx\Zout(s)$ (orange in \cref{fig:gap}b). It ranks $s$ by its estimated contribution to $\Zfn$, the \emph{rank score}
\begin{equation}
\mathrm{rank\ score}(s)=\Zin(s)\;\spr(s).\label{eq:score}
\end{equation}
At each position, \LN computes the rank score (\cref{eq:score}) of each new state, keeps the $\bfour$ four-index states and the $\btwo$ two-index states with the largest rank scores, and discards the rest, as \LinearPartition does.

\subsection{\texorpdfstring{Why \DPthree's partial stems are hard to score}{Why DP03's partial stems are hard to score}}\label{sec:beamdp}
A beam needs each state's outside weight, which is costly for two of \DPthree's three partial-stem states (\cref{tab:openidx}).

\begin{table}[t]%
\begin{minipage}[c]{0.40\linewidth}\centering\resizebox{\linewidth}{!}{\begin{tikzpicture}[x=1bp, y=-1bp]
\input{figs/dp03fig_defs}%
\input{figs/l2rfig_defs}%
\tikzset{oarm/.style={line width=1.6bp}, oopen/.style={line width=0.45bp, dash pattern=on 1.1bp off 0.7bp, <->, >={Stealth[length=1.8bp]}}}
\begin{scope}[shift={(-240.00,-150.40)}, scale=1.60]
\path[draw=stemB, line width=0.75bp] (150.00,119.00) .. controls (150.00,106.49) and (184.29,106.49) .. (184.29,119.00);
\path[draw=stemB, line width=0.75bp] (156.85,119.00) .. controls (156.85,110.84) and (177.43,110.84) .. (177.43,119.00);
\path[draw=black!75, line width=0.5bp] (150.00,119.00) -- (156.85,119.00);
\path[draw=black!75, line width=0.5bp] (177.43,119.00) -- (184.29,119.00);
\node[anchor=base west, inner sep=0pt, font=\scriptsize] at (148.51,125.89) {$x$};
\node[anchor=base west, inner sep=0pt, font=\scriptsize] at (155.58,125.89) {$c$};
\node[anchor=base west, inner sep=0pt, font=\scriptsize] at (176.31,125.89) {$f$};
\node[anchor=base west, inner sep=0pt, font=\scriptsize] at (183.01,125.89) {$j$};
\draw[stemA, oarm] (163.5,119.00) -- (170.8,119.00);
\draw[stemA, oopen] (157.6,116.6) -- (165.3,116.6);\draw[stemA, oopen] (169.0,116.6) -- (176.7,116.6);
\end{scope}
\node[anchor=base, inner sep=0pt, font=\footnotesize] at (27.36,14.00) {$\Qg_{x,c,f,j}$};
\node[anchor=base, inner sep=0pt, font=\footnotesize, text=black!70] at (27.36,60.00) {both ends open: $d,e$};
\begin{scope}[shift={(-169.00,-57.60)}, scale=1.60]
\path[draw=stemB, line width=0.75bp] (150.00,61.00) .. controls (150.00,48.49) and (184.29,48.49) .. (184.29,61.00);
\path[draw=stemB, line width=0.75bp] (161.76,61.00) -- (159.77,55.86) .. controls (166.39,53.57) and (177.43,55.28) .. (177.43,61.00);
\path[draw=black!75, line width=0.5bp] (150.00,61.00) -- (161.76,61.00);
\path[draw=black!75, line width=0.5bp] (177.43,61.00) -- (184.29,61.00);
\node[anchor=base west, inner sep=0pt, font=\scriptsize] at (148.51,67.90) {$x$};
\node[anchor=base west, inner sep=0pt, font=\scriptsize] at (160.62,67.90) {$d$};
\node[anchor=base west, inner sep=0pt, font=\scriptsize] at (176.31,67.90) {$f$};
\node[anchor=base west, inner sep=0pt, font=\scriptsize] at (183.01,67.90) {$j$};
\draw[stemA, oarm] (161.76,61.00) -- (170.0,61.00);
\draw[stemA, oopen] (168.2,58.6) -- (176.7,58.6);
\end{scope}
\node[anchor=base, inner sep=0pt, font=\footnotesize] at (98.36,14.00) {$\Qgl_{x,d,f,j}$};
\node[anchor=base, inner sep=0pt, font=\footnotesize, text=black!70] at (98.36,60.00) {one end open: $e$};
\begin{scope}[shift={(-95.60,-100.82)}, scale=1.60]
\path[draw=stemB, line width=0.75bp] (148.50,88.01) .. controls (148.50,75.49) and (182.79,75.49) .. (182.79,88.01);
\path[draw=stemB, line width=0.75bp] (160.26,88.01) -- (158.69,82.86) .. controls (162.81,81.56) and (168.49,81.56) .. (172.60,82.86) -- (171.04,88.01);
\path[draw=black!75, line width=0.5bp] (148.50,88.01) -- (160.26,88.01);
\path[draw=black!75, line width=0.5bp] (171.04,88.01) -- (182.79,88.01);
\node[anchor=base west, inner sep=0pt, font=\scriptsize] at (147.01,94.90) {$x$};
\node[anchor=base west, inner sep=0pt, font=\scriptsize] at (159.12,94.90) {$d$};
\node[anchor=base west, inner sep=0pt, font=\scriptsize] at (169.91,94.90) {$e$};
\node[anchor=base west, inner sep=0pt, font=\scriptsize] at (181.51,94.90) {$j$};
\draw[stemA, oarm] (160.26,88.01) -- (171.04,88.01);
\end{scope}
\node[anchor=base, inner sep=0pt, font=\footnotesize] at (169.36,14.00) {$\Qgr_{x,d,e,j}$};
\node[anchor=base, inner sep=0pt, font=\footnotesize, text=black!70] at (169.36,60.00) {no end open};
\end{tikzpicture}%
}\par\vspace{2pt}{\footnotesize\color{black!70}Stem B in Steps~2--3 of \cref{fig:pkassembly}; blue: A's arm; arrows: open ends.}\end{minipage}\hfill
\begin{minipage}[c]{0.57\linewidth}\centering\footnotesize\setlength{\tabcolsep}{2.6pt}\renewcommand{\arraystretch}{1.2}\captionsetup{font=footnotesize,skip=3pt}
\caption{\DPthree's three states for a partial stem (\cref{fig:states}c; steps of \cref{fig:pkassembly}). Open ends: the boundaries of the partner's arm that the state's indices leave free, as stem A (B's arm $x..c$) or as stem B (A's arm $d..e$). Terms: terms in the scan's estimate of its outside weight (\cref{app:termcounts}). Roles: the stems it can become.}\label{tab:openidx}
\begin{tabular}{@{}lccccc@{}}\toprule
 & & \multicolumn{2}{c}{Open ends} & & \\\cmidrule(lr){3-4}
State & Built at & as A & as B & Terms & Roles\\\midrule
$\Qg$, stem & -- & $x,c$ & $d,e$ & $\order(n^2)$ & A, B\\
$\Qgl$, $+1$ filler & Step~1 (A), 2 (B) & $c$ & $e$ & $\order(n)$ & A, B\\
$\QgrB$, $+2$ fillers & Step~3 (B) & \textbf{never} & none & 1 & B\\\bottomrule
\end{tabular}
\end{minipage}
\end{table}
\paragraph{Open partner boundaries and undecided roles.} \begin{edited}To score a partial stem, we must account for every placement of its crossing partner's arm inside its gap, and any boundary the state does not record must be summed over (\cref{app:dp03}). $\Qg$ leaves both ends of the partner's arm open and $\Qgl$ one (\cref{tab:openidx}, Open ends), so the scan's estimate has $\order(n^2)$ and $\order(n)$ terms (Terms). Neither state records its role (Roles): each can become stem A, completed by a partner B ahead of the scan ($\PKL$), or stem B, completed by a partner A behind it ($\PKR$; \cref{eq:outside}), and each case carries these sums. Only $\QgrB_{x,d,e,j}$ has neither problem: it records both ends of A's arm, and its partner A, $\Qgl_{i,x,d,e}$, is already built behind the scan, so its estimate is a single term with one completion, $\PKR$. \DPthree builds this state only for B: stem A stops at $\Qgl_{i,x,d,e}$ with B's end $c$ open (\cref{fig:gap}a), because attaching $\Ftwo$ to A would need $c$ before B exists.\end{edited}

\paragraph{\BeamDP.} We can still estimate these outside weights from \LN's pseudoknot-free passes and beam \DPthree's own recursion, keeping at each position the states with the largest rank score (\cref{eq:score}; App.~\ref{app:dp03outside}); we call this \BeamDP.
Its beam bounds how many states are kept, but not how many terms each outside weight has: on RNase P prefixes of 150 to 486\nt, its time grows as $n^{3.2}$ to $n^{3.8}$.
In \DPthree only $\QgrB$ records both; \cref{sec:pass3} builds A as $\Qgr$ too.%

\section{\texorpdfstring{\LRDP: \DPthree's Pseudoknot Rule, Left to Right}{LR-DP03: DP03's Pseudoknot Rule, Left to Right}}\label{sec:pass3}\label{sec:grammar}

\begin{figure}[!t]
\centering
\resizebox{0.89\linewidth}{!}{\input{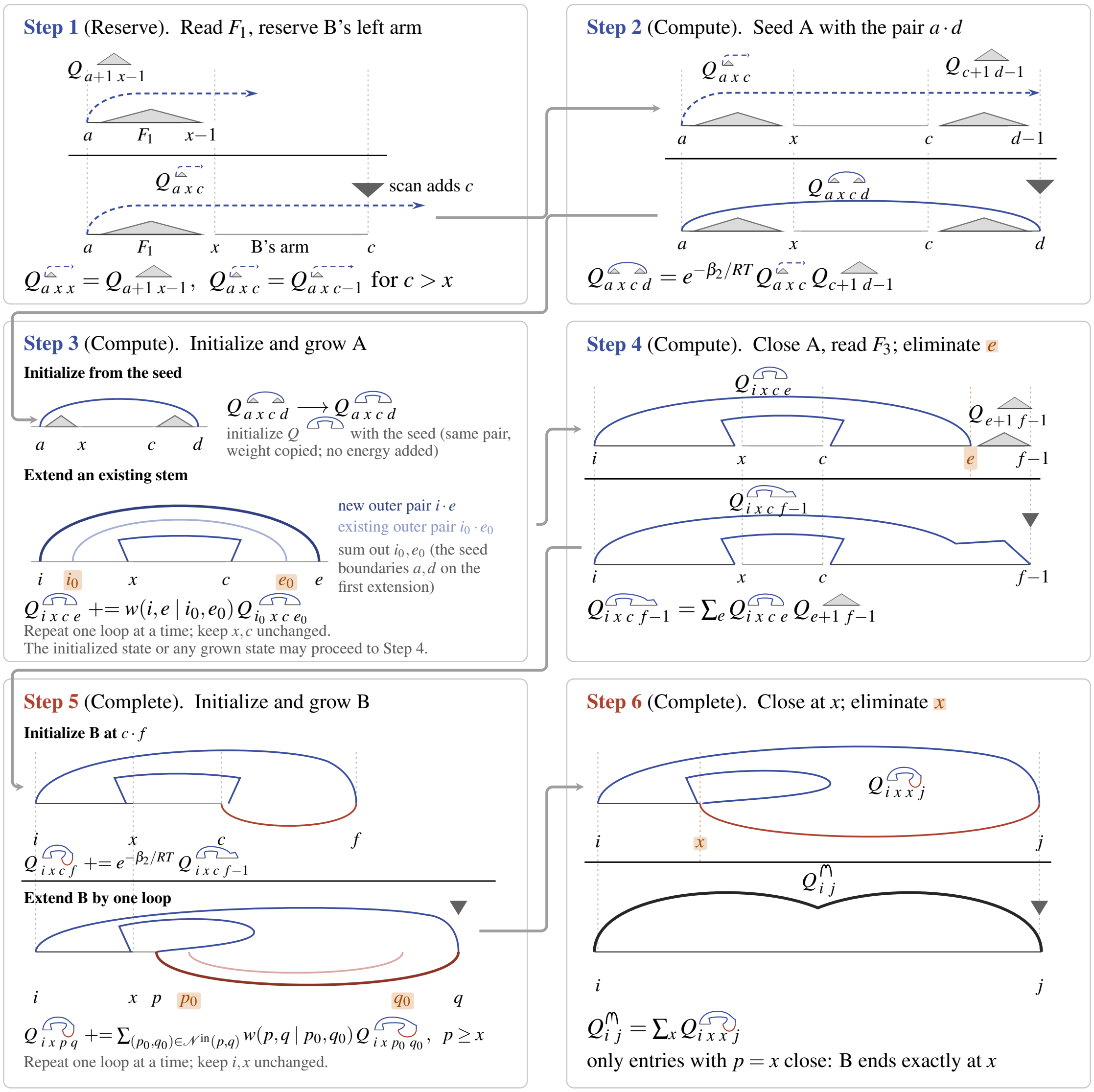}}
\caption{The L2R grammar, in the format of \cref{fig:pkassembly}; every item ends at the scan position (dotted line, triangle), and the gray segment $x..c$ is B's reserved left arm, still empty. In Step~3 ($i<i_0<x\le c<e_0<e$), the seed boundaries $a,d$ are not retained as separate indices after growth; contributions reaching the same $(i,x,c,e)$ are combined. The seed or any grown state may proceed to Step~4. Multiloop extensions use the helpers in \cref{fig:l2rmlsplit}. Step~5 initializes B at $\pair{c}{f}$ and extends it by one loop $\pair{p}{q}$ around $\pair{p_0}{q_0}$; on the first extension $p_0=c$ and $q_0=f$, so the orange $p_0,q_0$ are where $c,f$ are summed out (\cref{tab:elim}), and Step~6 closes it when $p=x$. Orange indices are summed out at that step (\cref{tab:elim}).}
\label{fig:l2rsteps}\label{fig:l2rstepsA}\label{fig:l2rstepsB}
\end{figure}

\paragraph{Speculation: put the partner's arm inside the state.}\label{sec:spec} We rewrite \DPthree's pseudoknot assembly so that A is built as $\Qgr_{i,x,c,e}$, with B's arm $x..c$ reserved before B exists. Eisner and Blatz~\cite{eisner2007} describe exact \emph{program transformations} for weighted dynamic programs, rewrites that compute the same value with different states or in a different order. One of them, \emph{speculation}, lets a rule start before one of its parts exists. Applied to the pseudoknot rule $\Qp$ (\cref{eq:Qp}) with stem B as the missing part, it has three phases, which the six steps of \cref{fig:l2rsteps} carry out: \textbf{Reserve} B's left arm $x..c$, by recording its ends $(x,c)$ in the state when the scan reads them, without B's weight, which is multiplied in when B is built (Step~1); \textbf{Compute} every factor that does not depend on B, which builds stem A as $\Qgr_{i,x,c,e}$ with $\Fone=\Qz_{a+1,x-1}$ and $\Ftwo=\Qz_{c+1,d-1}$ inside and then reads $\Fthree=\Qz_{e+1,f-1}$ (Steps~2--4); and \textbf{Complete} the rule by building B inside the state that holds A, closing the pseudoknot when B reaches the reserved $x$ (Steps~5--6). Each derivation of $\Qp$ is produced exactly once, as the part computed early times B attached later, so $\Zfn$ is unchanged; we call the result \LRDP (\cref{sec:exact}). Different reserved arms become separate states, so the beam chooses among them instead of summing.

The six steps below replace only \DPthree's assembly of $\Qp$; the rest of \DPthree is unchanged. Every state ends at the scan position $t$, and we name $t$ after the boundary it becomes in \cref{eq:Qp} ($x{-}1$, $c$, $d$, $e$, $f{-}1$, $j$), so the indices carry over from step to step. States that end at the same position are built in dependency order.

\paragraph{Step~1: read $\Fone$, reserve B's left arm.} The first filler $\Fone=\Qz_{a+1,x-1}$ (\cref{eq:Qz}) is an ordinary filler, which the \LRDP scan already builds; it needs no new state, and the base $a$ that waits for its partner $d$ is the base before it. B's left arm starts at $x$, $\Qlx_{a,x,x}=\Qz_{a+1,x-1}$, and $\Qlx_{a,x,c}=\Qlx_{a,x,c-1}$ for $c>x$ extends it by one base per position, so the scan position is the arm's right end $c$. $\Qlx_{a,x,c}$ records only the arm's ends and multiplies no weight; B's pairs come in Steps~5--6.

\paragraph{Step~2: seed A with the pair $\pair{a}{d}$.} When the scan reaches $d$ and $\pair{a}{d}$ can pair, the reserved arm has ended at some $c<d$, followed by the second filler $\Ftwo=\Qz_{c+1,d-1}$. For each such $c$, the state $\Qlx_{a,x,c}$ kept at $c$ joins $\Ftwo$, with \DPthree's penalty for A's inner pair: $\Qgrseed_{a,x,c,d}=e^{-\betatwo/\RT}\,\Qlx_{a,x,c}\,\Qz_{c+1,d-1}$. The seed is A's single pair $\pair{a}{d}$ around $\Fone$, B's arm and $\Ftwo$.

\paragraph{Step~3: grow A.} A $\Qgr$ state starts from a seed pair $\pair{a}{d}$ or extends an existing $\Qgr$ state by adding a new outer pair $\pair{i}{e}$ around its current pair $\pair{i_0}{e_0}$:
\begin{equation}\label{eq:Qgrgrow}
\begin{aligned}
&\text{initialize:} && \Qgr_{a,x,c,d}\mathrel{+}=\Qgrseed_{a,x,c,d},\\
&\text{extend:} && \Qgr_{i,x,c,e}\mathrel{+}=\textstyle\sum_{(i_0,e_0)\in\Nin(i,e)}\wloop{i}{e}{i_0}{e_0}\,\Qgr_{i_0,x,c,e_0}.
\end{aligned}
\end{equation}
Each extension keeps B's reserved arm $x..c$ unchanged. Stems with the same reserved arm that reach the same outer pair have their weights added into $\Qgr_{i,x,c,e}$, even if their seed pairs differ. Thus, the state stores $i,x,c,e$, but not the original seed boundaries $a,d$. Initialized from the seed $\Qgrseed_{a,x,c,d}$, the state $\Qgr_{a,x,c,d}$ may proceed directly to Step~4 or grow outward to $\Qgr_{i,x,c,e}$ before proceeding to Step~4. Multiloop growth uses the helpers in \cref{app:l2r}, \cref{fig:l2rmlsplit}.

\paragraph{Step~4: close A, read $\Fthree$.} After A's right end $e$, the scan reads the third filler $\Fthree=\Qz_{e+1,f-1}$ one base or one item per position, in the state $\Qgrz_{i,x,c,t}=\sum_e\Qgr_{i,x,c,e}\,\Qz_{e+1,t}$, which no longer records where A ended.

\paragraph{Step~5: seed and grow B.} B follows the same seed-and-grow procedure as A in Step~3. When the scan reaches $f$, the state $\Qgrz_{i,x,c,f-1}$ already contains A and all three fillers, including $\Fthree=\Qz_{e+1,f-1}$. Adding B's seed pair $\pair{c}{f}$, with \DPthree's inner-pair penalty, initializes $\Qpg$. We then grow B one loop at a time:
\begin{equation}\label{eq:Qpg}
\begin{aligned}
&\text{initialize at }\pair{c}{f}: && \Qpg_{i,x,c,f}\mathrel{+}=e^{-\betatwo/\RT}\,\Qgrz_{i,x,c,f-1},\\
&\text{grow to }\pair{p}{q}: && \Qpg_{i,x,p,q}\mathrel{+}=\textstyle\sum_{(p_0,q_0)\in\Nin(p,q)}\wloop{p}{q}{p_0}{q_0}\,\Qpg_{i,x,p_0,q_0},\quad p\ge x.
\end{aligned}
\end{equation}
A $\Qpg$ state starts from the seed pair $\pair{c}{f}$ or extends an existing $\Qpg$ state by adding a new outer pair $\pair{p}{q}$ around its current pair $\pair{p_0}{q_0}$. Each extension keeps $i,x$ unchanged. Stems with the same $i,x$ that reach the same outer pair have their weights added into $\Qpg_{i,x,p,q}$, even if their seed pairs differ. Thus, as in Step~3, the original seed boundaries $c,f$ are no longer stored separately. B grows only as far left as the reserved boundary $x$; when $p=x$, the state may proceed to Step~6, directly from its seed or after further growth. Multiloop extensions use the same helper construction as for A (\cref{app:l2r}, \cref{fig:l2rmlsplit}). From initialization onward, A and B are held together in one beam entry.

\paragraph{Step~6: close at $x$.} The pseudoknot closes when B's left end reaches the reserved $x$, at B's final right end $j$: $\Qp_{i,j}=\sum_x\Qpg_{i,x,x,j}$. Only entries with $p=x$ close, so closing checks that B ends exactly at $x$; summing over $x$ is the last elimination, and $\Qp$ keeps only the pseudoknot's ends $i$ and $j$. The six steps compute $\Qp$ exactly (\cref{sec:exact}, \cref{tab:elim}).

\section{\LRDP Is Exact and Counts Each Structure Once}\label{sec:exact}

\paragraph{Single counting in \LRDP.} \LRDP only regroups the sum \cref{eq:Qp}: each of its six inner boundaries is summed out exactly once, at a different step than in \DPthree (\cref{tab:elim}), so each structure in \DPthree's class keeps exactly one derivation, and without beams \LRDP computes \DPthree's $\Zfn$ in $\order(n^5)$ time and $\order(n^4)$ memory (\cref{app:l2rsingle}). Pruning only removes derivations, so the beamed $\Zfn$ never exceeds the exact one.

\paragraph{Numerical check.} Without beams, \LRDP reproduces the pseudoknot partition function of \NUPACK~3.0.6, an independent implementation of \DPthree: on 371 \PseudoBase sequences of 21 to 190\nt, the largest $|\Delta\log\Zfn|$ is $4.8\times10^{-12}$ and the largest pair-probability change is $4.7\times10^{-12}$, at floating-point rounding.%

\section{\texorpdfstring{Outside Estimates of the \LRDP States and the Three Passes}{Outside Estimates of the LR-DP03 States and the Three Passes}}\label{sec:overview}

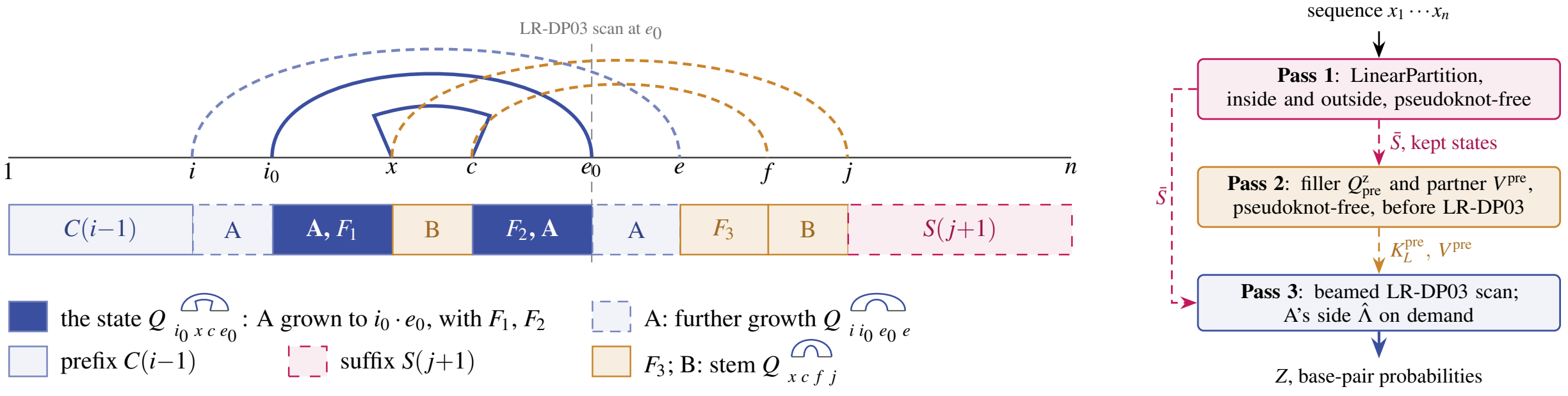
\begin{figure}[H]%
\begin{minipage}[b]{0.70\linewidth}\centering\resizebox{\linewidth}{!}{\providecolor{stemA}{HTML}{3B4FA0}
\providecolor{stemB}{HTML}{B5402E}
\providecolor{passtwo}{HTML}{C9852B}
\providecolor{passone}{HTML}{C2185B}
\begin{tikzpicture}[x=0.40cm, y=1cm, font=\scriptsize]
  \tikzset{
    seg/.style={draw=stemA!70, line width=0.45pt, minimum height=0.50cm, inner sep=0pt, anchor=south west},
    inside/.style={seg, fill=stemA, text=white, font=\scriptsize\bfseries},
    scan/.style={seg, fill=stemA!7, text=stemA!80!black},
    lazy/.style={seg, draw=stemA!70, densely dashed, fill=stemA!7, text=stemA!80!black},
    pre/.style={seg, draw=passtwo, fill=passtwo!14, text=passtwo!80!black},
    one/.style={seg, draw=passone, dashed, fill=passone!8, text=passone!80!black},
    ours/.style={line width=1.0pt},
    fut/.style={line width=0.8pt, dash pattern=on 2.5pt off 1.6pt},
  }
  \def\LRcell#1#2#3#4{\node[#1, minimum width={(#3-#2)*0.40cm}] at (#2,-0.98) {#4};}
  \def\LRarc#1#2#3#4{\draw[#1] (#2,0) .. controls (#2,#4) and (#3,#4) .. (#3,0);}
  \def\xi{4.6}\def\xp{6.6}\def\xa{7.8}\def\xx{9.6}\def\xc{11.6}\def\xd{13.2}\def\xq{14.6}\def\xe{16.8}
  \def\xf{19.0}\def\xj{21.0}\def\xn{26.6}
  \draw[black!75, line width=0.45pt] (0,0) -- (\xn,0);
  \draw[black!45, densely dashed] (\xq,-1.05) -- (\xq,1.14);%
  \node[above, inner sep=1pt, text=black!55, font=\tiny] at (\xq,1.14) {LR-DP03 scan at $e_0$};
  \LRarc{stemA!70, fut}{\xi}{\xe}{1.45}
  \LRarc{stemA, ours}{\xp}{\xq}{1.12}
  \draw[stemA, ours] (\xx,0) -- ({\xx-0.45},0.42) .. controls ({\xx+0.6},0.55) and ({\xc-0.6},0.55) .. ({\xc+0.45},0.42) -- (\xc,0);
  \LRarc{passtwo, fut}{\xx}{\xj}{1.30}
  \LRarc{passtwo, fut}{\xc}{\xf}{0.98}
  \foreach \x/\l in {0/1,\xi/i,\xp/i_0,\xx/x,\xc/c,\xq/e_0,\xe/e,\xf/f,\xj/j,\xn/n}
    \node[below, inner sep=1.2pt] at (\x,0) {$\l$};
  \LRcell{scan}{0}{\xi}{$\Cpref(i{-}1)$}
  \LRcell{lazy}{\xi}{\xp}{A}
  \LRcell{inside}{\xp}{\xx}{A, $\Fone$}
  \LRcell{pre}{\xx}{\xc}{B}
  \LRcell{inside}{\xc}{\xq}{$\Ftwo$, A}
  \LRcell{lazy}{\xq}{\xe}{A}
  \LRcell{pre}{\xe}{\xf}{$\Fthree$}
  \LRcell{pre}{\xf}{\xj}{B}
  \LRcell{one}{\xj}{\xn}{$\Sfull(j{+}1)$}
  \begin{scope}[shift={(0,-1.75)}]
    \node[inside, minimum width=0.38cm, minimum height=0.30cm] at (0,0) {};  \node[anchor=west] at (1.0,0.15) {the state $\Qgr_{i_0,x,c,e_0}$: A grown to $\pair{i_0}{e_0}$, with $\Fone$, $\Ftwo$};
    \node[scan,   minimum width=0.38cm, minimum height=0.30cm] at (0,-0.45) {}; \node[anchor=west] at (1.0,-0.30) {prefix $\Cpref(i{-}1)$};
    \node[lazy,   minimum width=0.38cm, minimum height=0.30cm] at (14.6,0) {}; \node[anchor=west] at (15.6,0.15) {A: further growth $\Qg_{i,i_0,e_0,e}$};
    \node[pre,    minimum width=0.38cm, minimum height=0.30cm] at (14.6,-0.45) {}; \node[anchor=west] at (15.6,-0.30) {$\Fthree$; B: stem $\Qg_{x,c,f,j}$};
    \node[one,    minimum width=0.38cm, minimum height=0.30cm] at (7.0,-0.45) {}; \node[anchor=west] at (8.0,-0.30) {suffix $\Sfull(j{+}1)$};
  \end{scope}
\end{tikzpicture}%
}\\[-3pt]{\small (a)}\end{minipage}\hfill
\begin{minipage}[b]{0.27\linewidth}\centering\resizebox{\linewidth}{!}{\providecolor{stemA}{HTML}{3B4FA0}
\providecolor{passtwo}{HTML}{C9852B}
\providecolor{passone}{HTML}{C2185B}
\begin{tikzpicture}[font=\scriptsize, >=Stealth,
    box/.style={draw, rounded corners=2pt, align=center, inner sep=3pt, text width=4.1cm, line width=0.6pt},
    rank/.style={->, densely dashed, line width=0.6pt},
    lab/.style={font=\scriptsize, align=left, inner sep=1.5pt}]
  \node (seq) {sequence $x_1\cdots x_n$};
  \node[box, draw=passone, fill=passone!8, below=0.32cm of seq] (p1)
    {\textbf{Pass 1}: \LinearPartition,\\inside and outside, pseudoknot-free};
  \node[box, draw=passtwo, fill=passtwo!14, below=0.55cm of p1] (p2)
    {\textbf{Pass 2}: filler $\Qzpre$ and partner $\hV$,\\pseudoknot-free, before \LRDP};
  \node[box, draw=stemA, fill=stemA!7, below=0.55cm of p2] (p3)
    {\textbf{Pass 3}: beamed \LRDP scan;\\A's side $\hLam$ on demand};
  \node[below=0.32cm of p3] (out) {$\Zfn$, base-pair probabilities};
  \draw[->, line width=0.6pt] (seq) -- (p1);
  \draw[rank, passone] (p1) -- node[lab, right, xshift=2pt, text=passone!85!black] {$\Sbar$, kept states} (p2);
  \draw[rank, passtwo] (p2) -- node[lab, right, xshift=2pt, text=passtwo!85!black] {$\KL,\ \hV$} (p3);
  \draw[rank, passone] (p1.west) -- ++(-0.3,0) |- node[lab, pos=0.25, left, text=passone!85!black] {$\Sbar$} (p3.west);
  \draw[->, line width=0.9pt, stemA] (p3) -- (out);
\end{tikzpicture}%
}\\[-3pt]{\small (b)}\end{minipage}
\caption{(a)~The exact outside weight of the growing state $\Qgr_{i_0,x,c,e_0}$ in the exterior loop (\cref{eq:keys}) at \LRDP scan position $e_0$ (dashed line), with the indices of \cref{eq:Qp}. The state holds A grown to $\pair{i_0}{e_0}$, with $\Fone$ and $\Ftwo$ merged into it and B's reserved arm $x..c$ as its empty gap (\cref{fig:states}d); A's final outer pair $\pair{i}{e}$ is not reached yet (dashed). Blue cells form $\Lambda(i_0,e_0)[e]$: the prefix and A's further growth $\Qg_{i,i_0,e_0,e}$ (\cref{eq:Vex}). Orange and pink cells form $\KLex(e;x,c)$: $\Fthree$, B's stem $\Qg_{x,c,f,j}$ on the reserved arm $x..c$, and the suffix (\cref{eq:Vex,eq:KLex}). (b)~The three passes; dashed: estimates that only rank states.}\label{fig:outsideLR}\label{fig:pipeline}\label{fig:passflow}
\end{figure}

\subsection{\texorpdfstring{Outside weights in the exterior loop}{Outside weights in the exterior loop}}\label{sec:beamscan}\label{sec:outsideweight}
\begin{edited}We write the exact outside weight of the hardest \LRDP state, the growing stem A, $\Qgr_{i_0,x,c,e_0}$ (Step~3 of \cref{fig:l2rsteps}), with the indices of \cref{eq:Qp}: A has grown from its seed pair $\pair{a}{d}$ to its current outer pair $\pair{i_0}{e_0}$ but has not reached its final outer pair $\pair{i}{e}$ (dashed in \cref{fig:outsideLR}a), and the right arm of its partner B is not yet fixed. By the two properties identified in \cref{sec:beamdp}, no sum over partner-arm placements remains, and the state has one completion, by B. In the exterior loop,
\begin{equation}
\Zout(\Qgr_{i_0,x,c,e_0})=\omegaone\sum_{e\ge e_0}\Lambda(i_0,e_0)[e]\;\KLex(e;x,c).\label{eq:keys}
\end{equation}
Both factors are \DPthree's gap state $\Qg$ (\cref{eq:Qg}), a stem's weight from inner to outer pair, closed on one side:
\begin{equation}
\Lambda(i_0,e_0)[e]=\sum_{i\le i_0}\Cpref(i{-}1)\,\Qg_{i,i_0,e_0,e},\qquad \Vex[c][f](x)=\sum_{j\ge f}\Qg_{x,c,f,j}\,\Sfull(j{+}1).\label{eq:Vex}
\end{equation}
$\Lambda$ closes A's stem on the left with the prefix and keeps A's right end $e$, which the \LRDP scan has not reached; $\Vex$ closes B's stem on the right with the suffix and keeps B's left end $x$, which must meet the reserved arm. They are the same stem weight, closed on opposite sides. With the third filler and B's inner-pair penalty, $\Vex$ gives the \emph{arm weight}
\begin{equation}
\KLex(e;x,c)=\sum_{f>e}\Qz_{e+1,f-1}\,e^{-\betatwo/\RT}\,\Vex[c][f](x).\label{eq:KLex}
\end{equation}
Here $\Cpref(i{-}1)=Q_{1,i-1}$ and $\Sfull(j{+}1)=Q_{j+1,n}$ are the prefix and suffix weights (\cref{sec:target}), $\omegaone$ is the pseudoknot initiation, and A's inner-pair penalty is inside the state. $\Qg_{i,i_0,e_0,e}$ includes the term with no further growth, $\pair{i}{e}=\pair{i_0}{e_0}$, in which A stops at once; the seed is the case $\pair{i_0}{e_0}=\pair{a}{d}$. The other two states need less: once A is closed, $\Lambda$ drops; once B is being built, A and all three fillers are inside the state, and only $\Vex$ and the prefix remain. \Cref{app:keys} gives all four.\end{edited}

\subsection{\texorpdfstring{Where the factors of the outside weight come from: the three passes}{Where the factors of the outside weight come from: the three passes}}\label{sec:supply}
The \LRDP scan scores the state $\Qgr_{i_0,x,c,e_0}$ at position $e_0$ after reading only $[1,e_0]$ (\cref{fig:outsideLR}a; \cref{eq:keys}). The prefix $\Cpref$ lies before $e_0$, and the scan computes it exactly over its kept states. The suffix $\Sfull$ lies after $e_0$ and is nested, so \LinearPartition estimates it before the scan (Pass~1, \cref{sec:pass1}). $\Fthree$ and partner B, in $\KLex$, lie after $e_0$ and cross the scored stem, so they need pseudoknot-free \emph{gap} states, which \LinearPartition does not build (Pass~2). A's stem $\Lambda$ spans $e_0$, and the beamed scan builds it on demand (Pass~3, \cref{sec:pass3beam}; \cref{fig:outsideLR}b).

\section{\texorpdfstring{Passes 1 and 2: The Pseudoknot-Free Estimates}{Passes 1 and 2: The Pseudoknot-Free Estimates}}\label{sec:pass1}\label{sec:pass2}

\paragraph{Pass~1} runs \LinearPartition~\cite{zhang2020linearpartition} on the pseudoknot-free part of \DPthree's model, in $\order(n\bone\log\bone)$ time and $\order(n\bone)$ memory: it scans over two-index nested states, keeps the $\bone$ best at each position by their prefix score (\cref{sec:target}), and runs the outside pass over the survivors. It supplies three things. (i)~The suffix $\Sbar(k)\approx\Sfull(k)=Q_{k,n}$, used inside the partner table $\hV$ (\cref{eq:V}). (ii)~The nested prefix $\Cpre(k)\approx Q_{1,k}$, used only to rank entries inside Pass~2; Pass~3 builds its own prefix $\Cpref$, pseudoknots included. (iii)~Its surviving states: Pass~2 builds B's stem in $\hV$, and Pass~3 extends A's stem $\Qg$ in $\hLam$, only over the pairs and multiloop states that Pass~1 keeps.

\paragraph{Pass~2}\label{sec:outside}\label{sec:partner} After Pass~1, we have nested suffix estimates but still need the crossing partner to score stem A. Pass~2 estimates the contribution of stem B, the third filler $\Fthree$, and the suffix beyond B (\cref{fig:outsideLR}); together, these estimate the arm weight $\KLex$ (\cref{eq:KLex}). Pass~2 builds two pseudoknot-free tables: the partner table answers how partner B and the suffix can complete A's pseudoknot, and the filler table supplies $\Fthree$ between A and B.
\begin{enumerate}[label=(\arabic*),leftmargin=*]%
\item \textbf{Partner table} $\hV[c][f](x)$, the estimate of B's side $\Vex$ (\cref{eq:Vex}): B's stem $\Qg_{x,c,f,j}$, pseudoknot-free, extended outward by the gap-state recursion (\cref{eq:Qg}) from its inner pair $\pair{c}{f}$ until its left end reaches $x$, the left end of B's arm reserved in A's gap; it is summed over B's right end $j$ and multiplied by the suffix after $j$:
\begin{equation}
\hV[c][f](x)=\ind{c=x}\,\Sbar(f{+}1)+\sum_{(c'',f'')\in\Nout(c,f),\;c''\ge x}\wloop{c''}{f''}{c}{f}\;\hV[c''][f''](x),\label{eq:V}
\end{equation}
where $\Nout(c,f)$ is the set of pairs one loop outside $\pair{c}{f}$. The first term stops B's stem: its left end has reached $x$, so $\pair{c}{f}$ is B's outer pair, $j=f$, and the suffix is $\Sbar(f{+}1)$. The second term extends B's stem by one loop to $\pair{c''}{f''}$, never past $x$; multiloop branches come from Pass~1. %
\item \textbf{Filler table} $\Qzpre(e{+}1,f{-}1)$: the filler $\Fthree$ between A's outer end $e$ and B's inner end $f$, computed by the filler recursion $\Qz$ (\cref{eq:Qz}) without its pseudoknot case.
\end{enumerate}
Attaching the filler to the partner gives the \emph{arm weight}, the estimate of $\KLex$ (\cref{eq:KLex}):
\begin{equation}
\KL(e;x,c)=\sum_{f>e}\Qzpre(e{+}1,f{-}1)\;e^{-\betatwo/\RT}\;\hV[c][f](x),\label{eq:KL}
\end{equation}
where $e^{-\betatwo/\RT}$ is the penalty for B's inner pair $\pair{c}{f}$; it is \cref{eq:KLex} with $\Qzpre$ and $\hV$ in place of $\Qz$ and $\Vex$.

\paragraph{Beaming Pass~2.} Both tables follow \LinearPartition's rule (\cref{sec:target}): two entries are compared only when, after multiplying each by Pass~1's nested prefix up to its left end, both cover the same part of the sequence. $\Qzpre$ keeps $\btwo$ entries per right end, and $\hV$ keeps $\bV$ entries per map (\cref{app:pass2}). For fixed $\bone$, $\btwo$ and $\bV$, Pass~2 is linear in $n$; unbeamed, it is exact.

\section{\texorpdfstring{Pass 3: Beamed \LRDP}{Pass 3: Beamed LR-DP03}}\label{sec:pass3beam}

\begin{edited}Pass~3 runs the \LRDP scan of \cref{sec:pass3} from left to right. It computes the outside estimate $\spr$ of every state by replacing each factor of its outside weight (\cref{eq:keys}; \cref{eq:keysall} for the other states) with an estimate: the prefix $\Cpref$ from the states it retains, $\KL$ for $\KLex$ and $\hV$ for $\Vex$ (\cref{sec:pass2}), and $\hLam$ for $\Lambda$.

\paragraph{A's side, on demand.} The outside weight has two sides: B's side, $\Vex$, which Pass~2 builds before the \LRDP scan as $\hV$, and A's side, $\Lambda$, which closes the scored stem with the prefix (\cref{eq:Vex}). Pass~3 builds A's side only when a candidate needs a score: it extends A's stem outward by the gap-state recursion (\cref{eq:Qg}), one loop at a time over the pairs that Pass~1 keeps, closes it with its own prefix $\Cpref$, and caches $\hLam$ with $K=32$ right ends per state. \Cref{sec:growth} explains why Pass~3, not Pass~2, builds it.

\paragraph{Beam.} At each position, Pass~3 keeps, for each table, the $\bfour$ four-index or $\btwo$ two-index states with the largest rank score $\Zin\cdot\spr$ (\cref{eq:score}). Only Pass~3 contributes to $\Zfn$: Passes~1 and~2 only rank states, kept states are combined by \LRDP's exact recursions, and the only approximation in $\Zfn$ is which states the beam keeps (\cref{app:keys}).\end{edited}

\section{Results}\label{sec:eval}

\subsection{Setup}
We use \DPthree's energy model: rna1999 parameters~\cite{mathews1999}, ``some'' dangles, interior loops capped at 30\nt, and \DPthree's pseudoknot penalties. Exact \DPthree is \NUPACK~3.0.6 with this model. By default, the beams are $\bone=200$ (Pass~1), $\btwo=100$ (two-index) and $\bfour=3000$ (four-index). We evaluate on \PseudoBase{} 2.0~\cite{vanbatenburg2000,taufer2009pseudobase} (445), \ArchiveII~\cite{sloma2016} (3{,}843) and Rous sarcoma virus genome prefixes up to 9{,}392\nt. Baselines are CParty~\cite{gray2025cparty}, \HotKnots~2.0~\cite{ren2005,andronescu2010}, \IPknot~\cite{sato2011} and \LinearPartition~\cite{zhang2020linearpartition} with \ThreshKnot ($\theta=0.3$). CParty and \HotKnots use the \DPnine parameters~\cite{andronescu2010}. We give CParty no input structure, so it chooses $G$ itself, as its lowest-energy hotspot (a single stable stem); its pair probabilities are pair frequencies in 1{,}000 stochastic samples, which we decode with \ThreshKnot. Times are wall-clock, on one AMD EPYC 9474F node.

\subsection{Efficiency and scalability}
\label{sec:eff}

\begin{figure}[t]
\centering
\begin{minipage}[b]{0.245\linewidth}\centering\includegraphics[width=\linewidth]{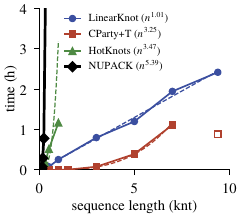}\\[-5pt]{\small (a)}\end{minipage}\hfill
\begin{minipage}[b]{0.245\linewidth}\centering\includegraphics[width=\linewidth]{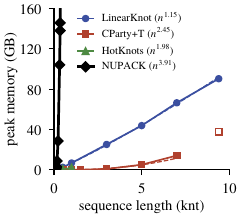}\\[-5pt]{\small (b)}\end{minipage}\hfill
\begin{minipage}[b]{0.245\linewidth}\centering\includegraphics[width=\linewidth]{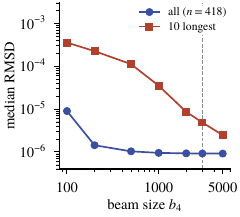}\\[-5pt]{\small (c)}\end{minipage}\hfill
\begin{minipage}[b]{0.245\linewidth}\centering\includegraphics[width=\linewidth]{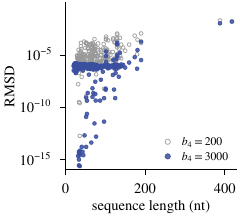}\\[-5pt]{\small (d)}\end{minipage}
\caption{(a)~Time and (b)~peak memory against sequence length.
Inputs are prefixes of the Rous sarcoma virus genome up to the full 9{,}392\nt, run on an AMD EPYC 9474F node; open square: CParty did not finish.
NUPACK is measured on \PseudoBase and \ArchiveII sequences of 48 to 418\nt; its longest runs are off scale (11.5 and 18.8\,h, 210 and 283\,GB at 388 and 418\nt). Dashed: fits.
(c)~Median RMSD of \LN base-pair probabilities against exact \DPthree on 418 \PseudoBase sequences, against beam size $\bfour$ ($\btwo=1000$, $\bV=256$; dashed: default).
(d)~Per-sequence RMSD against length.
}
\label{fig:results}\label{fig:scaling}\label{fig:rmsd}\label{fig:rmsdlen}
\end{figure}

\Cref{fig:results}a,b compare \LN with exact \DPthree, run as \NUPACK, and with two faster pseudoknot methods, CParty with \ThreshKnot and \HotKnots.
On prefixes of the Rous sarcoma virus genome from 1{,}000\nt to the full 9{,}392\nt, the time of \LN grows as $n^{1.01}$ and its memory as $n^{1.15}$; the full genome takes 2.4\,h and 90.5\,GB.
\NUPACK grows as $n^{5.39}$ in time and $n^{3.91}$ in memory over 233 to 418\nt; at 388\nt it takes 11.5\,h and 210\,GB, more than \LN needs for the whole genome, and on the 418\nt PKB10019, 18.8\,h and 283\,GB.
CParty is faster than \LN and uses less memory up to 7{,}000\nt, but grows as $n^{3.25}$ and did not finish the full genome, nor did \HotKnots ($n^{3.47}$) at 1{,}500\nt.

\subsection{Approximation quality against exact \DPthree}
\label{sec:approx}\label{sec:beamsize}

We measure the error of \LN's base-pair probabilities by their RMSD against exact \DPthree over all pairs $i<j$, on the 418 \PseudoBase sequences whose bases are all A, C, G or U (\cref{fig:results}c,d).
At the default $\bfour=3000$, the median RMSD is $9.2\times10^{-7}$ (mean $6.8\times10^{-5}$); 409 of the 418 sequences are below $10^{-5}$ and 416 below $10^{-3}$.
Most sequences converge at small beams: the median falls from $9.0\times10^{-6}$ at $\bfour=100$ to $1.4\times10^{-6}$ at 200 and then barely changes.
For the 10 longest sequences (145 to 418\nt), it keeps falling, from $3.6\times10^{-4}$ at $\bfour=100$ to $2.5\times10^{-6}$ at 5000.

The two longest \PseudoBase sequences (PKB10019, 418\nt; PKB10003, 388\nt) have the largest RMSD at $\bfour = 3000$ ($1.7 \times 10^{-2}$ and $1.1 \times 10^{-2}$).
Following \LinearPartition~\cite{zhang2020linearpartition}, we ask whether these deviations from the exact ensemble move probability toward or away from the reference structure (\cref{fig:pkb10019}b shows PKB10019).
Among pairs where \LN assigns at least 0.2 more probability than exact \DPthree, 25 of 40 and 23 of 39 are in the reference structure; among pairs where exact \DPthree is higher by at least 0.2, only 11 of 44 and 7 of 38 are.
Consequently, the \ThreshKnot structure from \LN is more accurate than that from the exact ensemble on both sequences (F1 0.848 vs.\ 0.764 and 0.757 vs.\ 0.661), and on PKB10003 it also recovers more crossing pairs (crossing F1 0.537 vs.\ 0.359).

\subsection{Accuracy of downstream predictions}
\label{sec:downstream}

\begin{figure}[t]
\centering
\newsavebox{\avgaccbox}%
\sbox{\avgaccbox}{\footnotesize\setlength{\tabcolsep}{2.4pt}%
\begin{tabular}[b]{@{}lcccc@{}}
\toprule
& \multicolumn{2}{c}{\PseudoBase, seq.\ avg.} & \multicolumn{2}{c}{\ArchiveII, family avg.}\\
\cmidrule(lr){2-3}\cmidrule(l){4-5}
Method & All pairs & Crossing & All pairs & Crossing\\
& P\,/\,R\,/\,F1 & P\,/\,R\,/\,F1 & P\,/\,R\,/\,F1 & P\,/\,R\,/\,F1\\
\midrule
\LN & 75.3\,/\,\textbf{78.9}\,/\,\textbf{76.4} & 76.1\,/\,\textbf{69.1}\,/\,\textbf{67.1} & 44.0\,/\,51.7\,/\,47.1 & \phantom{0}5.9\,/\,\textbf{20.6}\,/\,\textbf{\phantom{0}7.0}\\
\midrule
\LinearPartition-V + T & 65.2\,/\,55.1\,/\,58.7 & 37.3\,/\,\phantom{0}2.8\,/\,\phantom{0}3.0 & 47.8\,/\,\textbf{56.4}\,/\,51.3 & \phantom{0}3.9\,/\,\phantom{0}2.5\,/\,\phantom{0}1.5\\
\LinearPartition-C + T & \textbf{77.3}\,/\,54.1\,/\,61.4 & 60.9\,/\,\phantom{0}3.8\,/\,\phantom{0}4.2 & \textbf{59.3}\,/\,51.8\,/\,\textbf{54.3} & \phantom{0}7.4\,/\,\phantom{0}1.4\,/\,\phantom{0}1.1\\
\IPknot & 76.5\,/\,64.8\,/\,68.9 & 62.3\,/\,30.5\,/\,32.0 & 54.2\,/\,55.5\,/\,54.2 & \phantom{0}6.7\,/\,13.7\,/\,\phantom{0}6.2\\
\IPknot (RNAfold) & 70.3\,/\,59.7\,/\,63.4 & 54.4\,/\,25.3\,/\,26.9 & 51.7\,/\,51.9\,/\,51.2 & \textbf{\phantom{0}7.7}\,/\,13.4\,/\,\phantom{0}6.5\\
CParty + T & 71.7\,/\,71.8\,/\,71.0 & 75.7\,/\,52.9\,/\,51.8 & 50.9\,/\,52.2\,/\,51.1 & \phantom{0}6.6\,/\,\phantom{0}2.6\,/\,\phantom{0}1.8\\
\HotKnots 2.0 & 75.1\,/\,73.1\,/\,73.1 & \textbf{76.8}\,/\,49.1\,/\,47.7 & -- & --\\
\bottomrule
\end{tabular}}%
\newsavebox{\pkbbox}%
\sbox{\pkbbox}{\includegraphics[height=\dimexpr\ht\avgaccbox+\dp\avgaccbox\relax]{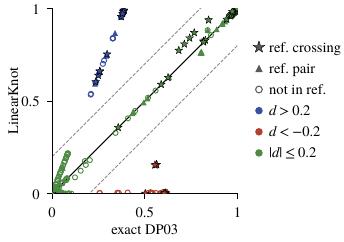}}%
\usebox{\avgaccbox}\hfill\usebox{\pkbbox}\\[-2pt]%
{\footnotesize\makebox[\wd\avgaccbox]{(a)}\hfill\makebox[\wd\pkbbox]{(b)}}
\caption{(a)~Accuracy (percent) averaged over sequences on \PseudoBase (445) and over families on \ArchiveII (3{,}843 sequences); +\,T: with \ThreshKnot.
(b)~Pair probabilities of \LN against exact \DPthree on PKB10019 (418\nt); colour gives $d$, the \LN probability minus the exact one.
}
\label{fig:avgacc}\label{fig:pkb10019}
\end{figure}

On all 445 \PseudoBase sequences, \LN has the highest F1 of all methods, 76.4 over pairs and 67.1 over crossing pairs, where a crossing pair must be crossing in both the reference and the prediction (\cref{fig:avgacc}a; definitions in App.~\ref{app:metrics}).
The next best are \HotKnots over pairs (73.1) and CParty with \ThreshKnot over crossing pairs (51.8).
The crossing score separates the methods most: pseudoknot-free \LinearPartition with \ThreshKnot recovers almost no crossing pairs as crossing (recall 2.8 and 3.8), \IPknot, \HotKnots and CParty recover 25.3 to 52.9, and \LN 69.1.
On \ArchiveII, \LN is less accurate over all pairs (F1 47.1), but it does best on crossing pairs.
Crossing F1 is low for every method there (at most 7.0), but \LN has the highest, and its crossing recall (20.6) is 1.5 times \IPknot's (13.7).

\subsection{\texorpdfstring{Case study: the Rous sarcoma virus frameshift element}{Case study: the Rous sarcoma virus frameshift element}}
\label{sec:rsv}
\begin{figure}[t]
\centering
\includegraphics[width=\linewidth]{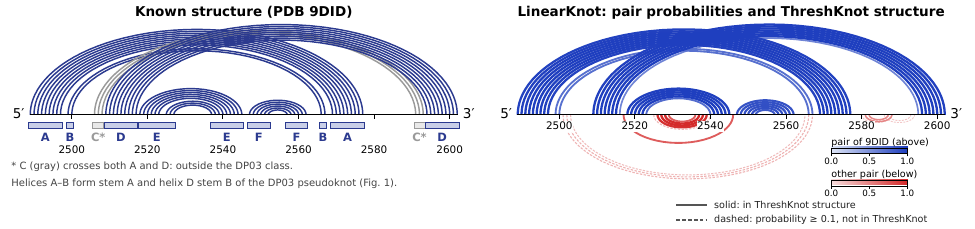}\vspace{-6pt}%
\caption{The pseudoknot of the Rous sarcoma virus \emph{gag-pol} frameshift element (NC\_001407.1, Prague C, positions 2489--2602, 114\nt; the slippery site and spacer are excluded). Left: the known structure 9DID~\cite{jones2025rsv}, one arc per base pair; the bars under the axis mark the two arms of each helix A--F. Helix C (gray) lies outside \DPthree's class. Right: \LN's base-pair probabilities ($\ge 0.1$) for the same 114\nt. Pairs of 9DID are drawn above the axis (blue) and all other pairs below (red); darker means more probable. Solid arcs are the pairs \ThreshKnot selects ($\theta=0.3$); dashed arcs have probability $\ge 0.1$ but are not selected.}
\label{fig:rsv}
\end{figure}
The \emph{gag-pol} frameshift element of Rous sarcoma virus~\cite{jones2025rsv} is shown in \cref{fig:rsv}: helices A--B and D form an H-type pseudoknot, with hairpins E and F inside it, and helix C lies outside \DPthree's class.
Folded alone (the 114\nt pseudoknot, 2489--2602), \LN recovers the pseudoknot: all 9 pairs of A and 8 of the 9 pairs of D have probability at least 0.5 (summed probabilities 8.94 and 7.57), the two pairs of B have 0.70 and 0.71, and all 6 pairs of F are predicted.
The inner three pairs of E occur shifted by one base on the $5'$ arm.
\LN's \ThreshKnot structure has the highest F1 against 9DID of all methods, 0.82, against 0.79 for the next best, \HotKnots (all methods in \cref{app:rsv}).

\section{Discussion}\label{sec:discussion}

\paragraph{Summary.} \LN approximates the \DPthree partition function in a left-to-right scan guided by two pseudoknot-free passes.
In \LN, the crossing partner of a stem and the stem's role become part of the state; in the original \DPthree formulation, they would be hidden in a sum. The beam can therefore choose them instead of summing over them.

\paragraph{Extensions and limitations.} Making arm choices explicit has a cost: an arm the beam discards early cannot be recovered later.  The outside estimate places every pseudoknot in the exterior loop and uses pseudoknot-free suffixes and partners, so it can misrank pseudoknots inside multiloops or fillers. Stages outside the scan are not yet beam-bounded, which keeps end-to-end memory marginally above linear ($n^{1.15}$). \DPthree's class itself excludes some pseudoknots, such as helix C of the Rous sarcoma virus element and kissing hairpins. A fast partition function makes it possible to refit \DPthree's pseudoknot parameters on large data, as Andronescu, Pop and Condon did with smaller data~\cite{andronescu2010}, and the same idea should extend to broader classes such as Rivas--Eddy's, which covers kissing hairpins.

\clearpage %
\bibliographystyle{abbrvnat}
\bibliography{references}

\ifdefined\submission\else
\clearpage
\appendix

\section{\DPthree's outside weight and folded evaluation}%
\label{app:dp03}
\paragraph{\DPthree's outside weight.} Dirks and Pierce~\cite{dirks2004} compute the outside weight of every \DPthree entry top-down from $Q_{1,n}$: each rule $\text{parent}=\sum\prod\text{factors}$ passes to every factor the parent's outside weight times the other factors of its term. They report it normalized, as the probability $\Zin(s)\,\Zout(s)/\Zfn$ that entry $s$ is used. A gap state $\Qg$ is a factor in exactly two rules: its own growth (\cref{eq:Qg}) and the assembly of the pseudoknot rule $\Qp$ (Steps~1--2 of \cref{fig:pkassembly}). Collecting the outside weight it receives from both rules gives \cref{eq:outside} below, which computes $\Zout$; \BeamDP (\cref{sec:beamdp}) ranks \DPthree's states by it.

\paragraph{The outside weight of a \DPthree gap state.}\label{app:dp03outside} Take a gap state $\Qg_{p,u,v,q}$ (\cref{fig:gap}) whose outer pair $\pair{p}{q}$ ends at the scan position $q$. To score it, we need everything that completes this partial stem. A partial stem must first grow to its final outer pair and then join a crossing partner as either stem A or stem B of \cref{fig:pkassembly}. Its outside weight sums over all possible growth paths and both roles:
\begin{equation}
\Zout(\Qg_{p,u,v,q})=\sum_{p'\le p,\;q'\ge q}\Gtab(p,q)[p',q']\,\Big(\PKL(p',q',u,v)+\PKR(p',q',u,v)\Big).\label{eq:outside}
\end{equation}
Here, $\Gtab(p,q)[p',q']$ accounts for growth from the current outer pair $\pair{p}{q}$ to the final pair $\pair{p'}{q'}$, including no growth. The terms $\PKL$ and $\PKR$ account for completing the finished stem as A or B, respectively; \cref{fig:outsideDP03} draws one completion of each. \Cref{eq:outside}, which computes $\Zout$, is \DPthree's own outside weight of the stem, and it is exact when the completion terms include every context allowed by \DPthree: the exterior loop, a multiloop, or a filler of another pseudoknot. Below, we derive the exterior-loop terms and use them in each context.

\paragraph{Notation of the generic stem.} As in the inside--outside algorithm, where an item's outside weight sums over its roles, we keep the scored stem's own indices and give each role's partner its own letters. \Cref{fig:htype} names the eight boundaries of a pseudoknot, $i,a,x,c,d,e,f,j$. The scored stem may still become A or B, so \cref{eq:outside} writes it with generic letters instead: $\pair{p}{q}$ and $\pair{u}{v}$ for its current outer and inner pairs, and $\pair{p'}{q'}$ for its final outer pair. If it becomes stem A, $\pair{u}{v}$ is A's inner pair $\pair{a}{d}$ and $\pair{p'}{q'}$ is A's outer pair $\pair{i}{e}$; if it becomes stem B, they are B's $\pair{c}{f}$ and $\pair{x}{j}$. The partner and the fillers keep the letters of \cref{fig:htype}a, as drawn in \cref{fig:outsideDP03}. The \LRDP states fix every role (\cref{sec:pass3}), so the main text uses only these letters.

\begin{figure}[ht]
\centering
\providecolor{stemA}{HTML}{3B4FA0}
\providecolor{stemB}{HTML}{B5402E}
\providecolor{passtwo}{HTML}{C9852B}
\providecolor{passone}{HTML}{C2185B}
\begin{tikzpicture}[x=0.40cm, y=1cm, font=\scriptsize]
  \tikzset{
    seg/.style={draw=stemA!70, line width=0.45pt, minimum height=0.36cm, inner sep=0pt, anchor=south west},
    inside/.style={seg, fill=stemA, text=white, font=\scriptsize\bfseries},
    scan/.style={seg, fill=stemA!7, text=stemA!80!black},
    lazy/.style={seg, draw=stemA!70, densely dashed, fill=stemA!7, text=stemA!80!black},
    pre/.style={seg, draw=passtwo, fill=passtwo!14, text=passtwo!80!black},
    one/.style={seg, draw=passone, dashed, fill=passone!8, text=passone!80!black},
    ours/.style={line width=1.0pt},
    fut/.style={line width=0.8pt, dash pattern=on 2.5pt off 1.6pt},
    built/.style={line width=0.8pt},
  }
  \def\LRcell#1#2#3#4{\node[#1, minimum width={(#3-#2)*0.40cm}] at (#2,-0.78) {#4};}
  \def\LRarc#1#2#3#4{\draw[#1] (#2,0) .. controls (#2,#4) and (#3,#4) .. (#3,0);}

  \begin{scope}[shift={(0,2.05)}]
    \node[inside, minimum width=0.38cm] at (0,0) {};   \node[anchor=west] at (1.0,0.18) {inside: the state};
    \node[scan, minimum width=0.38cm] at (6.2,0) {};   \node[anchor=west] at (7.2,0.18) {the scan (Pass 3)};
    \node[lazy, minimum width=0.38cm] at (13.4,0) {};  \node[anchor=west] at (14.4,0.18) {growth, on demand (Pass 3)};
    \node[pre, minimum width=0.38cm] at (25.0,0) {};   \node[anchor=west] at (26.0,0.18) {Pass 2 tables};
    \node[one, minimum width=0.38cm] at (32.0,0) {};   \node[anchor=west] at (33.0,0.18) {Pass 1 (\LinearPartition)};
  \end{scope}

  \begin{scope}
    \def\xpp{5.5}\def\xp{6.5}\def\xu{7.5}\def\xb{8.8}\def\xc{9.8}\def\xv{11.6}\def\xq{12.6}
    \def\xqq{13.6}\def\xf{15.2}\def\xj{16.2}\def\xn{19.5}
    \node[anchor=west, font=\footnotesize] at (0,1.62) {(a)~$\PKL$: the scored stem is A; partner B's right arm lies beyond the scan};
    \draw[black!75, line width=0.45pt] (0,0) -- (\xn,0);
    \draw[black!45, densely dashed] (\xq,-0.95) -- (\xq,1.38);
    \LRarc{stemA!60, fut}{\xpp}{\xqq}{1.45}               %
    \LRarc{stemA, ours}{\xp}{\xq}{1.2}                    %
    \LRarc{stemA, ours}{\xu}{\xv}{0.82}                   %
    \LRarc{passtwo, fut}{\xb}{\xj}{1.3}                   %
    \LRarc{passtwo, fut}{\xc}{\xf}{0.98}                  %
    \foreach \x/\l in {0/1,\xpp/p',\xp/p,\xu/u,\xb/x,\xc/c,\xv/v,\xq/q,\xqq/q',\xf/f,\xj/j,\xn/n}
      \node[below, inner sep=1.2pt] at (\x,0) {$\l$};
    \LRcell{scan}{0}{\xpp}{$\Cpref(p'{-}1)$}
    \LRcell{lazy}{\xpp}{\xp}{$\hG$}
    \LRcell{inside}{\xp}{\xu}{}
    \LRcell{scan}{\xu}{\xb}{$\Fone$}
    \LRcell{pre}{\xb}{\xc}{$\hV$}
    \LRcell{scan}{\xc}{\xv}{$\Ftwo$}
    \LRcell{inside}{\xv}{\xq}{}
    \LRcell{lazy}{\xq}{\xqq}{$\hG$}
    \LRcell{pre}{\xqq}{\xf}{$\Fthree$}
    \LRcell{pre}{\xf}{\xj}{$\hV$}
    \LRcell{one}{\xj}{\xn}{$\Sbar$, in $\hV$}
  \end{scope}

  \begin{scope}[shift={(20.9,0)}]
    \def\xi{1.4}\def\xa{2.4}\def\xpp{5.5}\def\xp{6.5}\def\xu{7.5}\def\xd{8.7}\def\xe{9.7}
    \def\xv{11.6}\def\xq{12.6}\def\xqq{13.6}\def\xn{19.5}
    \node[anchor=west, font=\footnotesize] at (0,1.62) {(b)~$\PKR$: the scored stem is B; partner A is behind the scan};
    \draw[black!75, line width=0.45pt] (0,0) -- (\xn,0);
    \draw[black!45, densely dashed] (\xq,-0.95) -- (\xq,1.38);
    \LRarc{stemA!60, fut}{\xpp}{\xqq}{1.45}               %
    \LRarc{stemA, ours}{\xp}{\xq}{1.2}                    %
    \LRarc{stemA, ours}{\xu}{\xv}{0.82}                   %
    \LRarc{stemA!38, built}{\xi}{\xe}{1.3}                %
    \LRarc{stemA!38, built}{\xa}{\xd}{0.98}               %
    \foreach \x/\l in {0/1,\xi/i,\xa/a,\xpp/p',\xp/p,\xu/u,\xd/d,\xe/e,\xv/v,\xq/q,\xqq/q',\xn/n}
      \node[below, inner sep=1.2pt] at (\x,0) {$\l$};
    \LRcell{scan}{0}{\xi}{$\Cpref$}
    \LRcell{scan}{\xi}{\xa}{A}
    \LRcell{scan}{\xa}{\xpp}{$\Fone$}
    \LRcell{lazy}{\xpp}{\xp}{$\hG$}
    \LRcell{inside}{\xp}{\xu}{}
    \LRcell{scan}{\xu}{\xd}{$\Ftwo$}
    \LRcell{scan}{\xd}{\xe}{A}
    \LRcell{scan}{\xe}{\xv}{$\Fthree$}
    \LRcell{inside}{\xv}{\xq}{}
    \LRcell{lazy}{\xq}{\xqq}{$\hG$}
    \LRcell{one}{\xqq}{\xn}{suffix $\Sbar(q'{+}1)$}
  \end{scope}
  \node[below, inner sep=1pt, text=black!55] at (12.6,-0.95) {scan};
  \node[below, inner sep=1pt, text=black!55] at (20.9+12.6,-0.95) {scan};
\end{tikzpicture}%
\caption{Two ways to complete the partial stem $\Qg_{p,u,v,q}$ at scan position $q$ (dashed line): the two terms of \cref{eq:outside}. After growing (dashed blue, $\hG$) from $\pair{p}{q}$ to a final outer pair $\pair{p'}{q'}$, the stem joins a crossing partner as (a)~stem A, the term $\PKL$ (\cref{eq:PKL}), or (b)~stem B, the term $\PKR$ (\cref{eq:PKR}); one completion of each is drawn. The state stores $p,u,v,q$; $p',q'$ and the partner's boundaries are summed over. In (a), partner B (orange) has its right arm beyond the scan; in (b), partner A lies before the scan, which has built it. The bar under each panel labels every interval by the source that would supply it to a scan over \DPthree's states (\BeamDP); colors as in \cref{fig:outsideLR}a. Other letters are those of \cref{fig:htype}a.}
\label{fig:outsideDP03}
\end{figure}
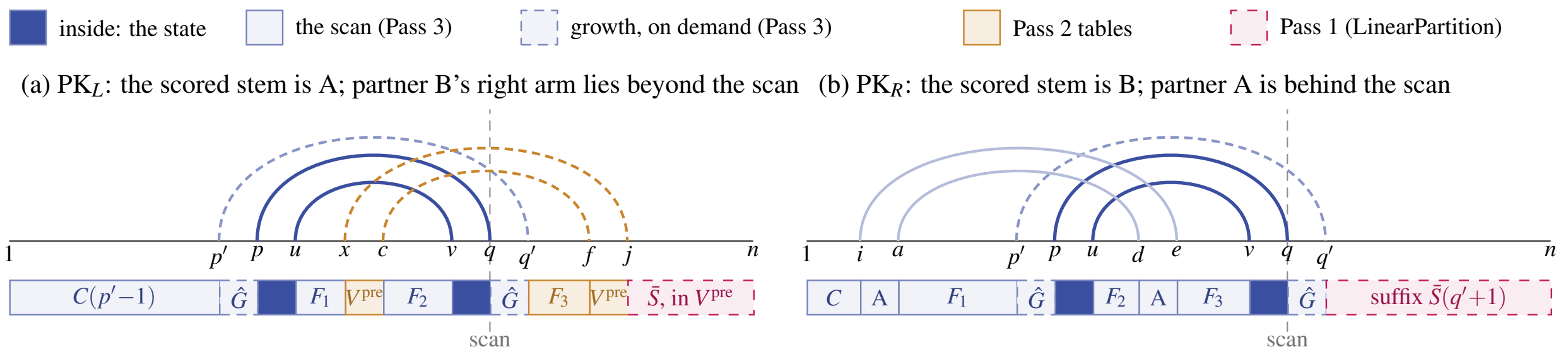

\paragraph{Completion as stem A} (\cref{fig:outsideDP03}a). If the scored stem becomes A, its completion term in \cref{eq:outside} is $\PKL(p',q',u,v)$. Fix the boundaries $x,c$ of B's left arm inside the stem's gap. This divides the gap into the first filler $\Fone$, B's left arm $x..c$, and the second filler $\Ftwo$, with filler weights $\Qz_{u+1,x-1}$ and $\Qz_{c+1,v-1}$. The completion weight $\KLex$ accounts for stem B, the third filler $\Fthree$, and the suffix beyond B. Multiplying by the prefix and the required penalties, then summing over $x,c$, gives
\begin{equation}
\PKL(p',q',u,v)\approx\omegaone\,e^{-\betatwo/\RT}\,\Cpref(p'{-}1)\sum_{u<x\le c<v}\Qz_{u+1,x-1}\;\KLex(q';x,c)\;\Qz_{c+1,v-1},\label{eq:PKL}
\end{equation}
where $\Cpref(p'{-}1)$ is the prefix before the stem, $\omegaone$ the pseudoknot initiation, $e^{-\betatwo/\RT}$ the penalty for the stem's inner pair $\pair{u}{v}$, and $\KLex(q';x,c)=\sum_{f>q'}\Qz_{q'+1,f-1}\,e^{-\betatwo/\RT}\,\Vex[c][f](x)$ the completion by B (cells $\Fthree$, $\hV$ and $\Sbar$ of \cref{fig:outsideDP03}a): the third filler $\Fthree=[q'{+}1,f{-}1]$, B's inner-pair penalty, and the \emph{partner weight} $\Vex[c][f](x)$, which is B grown outward by the gap-state recursion $\Qg$ (\cref{eq:Qg}) from its inner pair $\pair{c}{f}$ until its left end reaches $x$, summed over its right end $j$ and multiplied by the suffix $\Sfull(j{+}1)$. B's arm $x..c$ lies in the stem's gap, but its pairs are counted in $\Vex$.

\paragraph{Completion as stem B} (\cref{fig:outsideDP03}b). If the scored stem becomes B, its completion term is $\PKR(p',q',u,v)$. Fix the boundaries $d,e$ of A's right arm inside the stem's gap. This divides the gap into the second filler $\Ftwo$, A's right arm $d..e$, and the third filler $\Fthree$, with filler weights $\Qz_{u+1,d-1}$ and $\Qz_{e+1,v-1}$. The completion weight $\KR$ accounts for stem A, the first filler $\Fone$, and the prefix before A. Multiplying by the suffix and the required penalty, then summing over $d,e$, gives
\begin{equation}
\PKR(p',q',u,v)\approx\omegaone\,e^{-\betatwo/\RT}\,\Sfull(q'{+}1)\sum_{u<d\le e<v}\Qz_{u+1,d-1}\;\KR(p';d,e)\;\Qz_{e+1,v-1},\label{eq:PKR}
\end{equation}
where $\Sfull(q'{+}1)$ is the suffix after the stem, $\omegaone$ the pseudoknot initiation, $e^{-\betatwo/\RT}$ the penalty for the stem's inner pair $\pair{u}{v}$, and $\KR(p';d,e)=\sum_{i}\Cpref(i{-}1)\,\Qgl_{i,p',d,e}$ the prefix times partner A with $\Fone=[a{+}1,p'{-}1]$ attached (cells $\Cpref$, A and $\Fone$ of \cref{fig:outsideDP03}b), which is \DPthree's $\Qgl$ from Step~1 of \cref{fig:pkassembly}. For B's last state $\Qgr_{x,d,e,j}$ (\cref{tab:openidx}), both fillers are inside and the gap is exactly A's arm $d..e$, so the sum over $d,e$ disappears and its outside weight is the single term $\omegaone\,\Sfull(j{+}1)\,\KR(x;d,e)$.

\paragraph{Term counts of \cref{tab:openidx}.}\label{app:termcounts} \begin{edited}We count the terms of the scan's estimate of a stem's outside weight, per final outer pair, counting the completion by the partner for one arm placement as one term. Take stem A. Besides its growth edges (\cref{eq:Qg}), $\Qg_{i,a,d,e}$ has one Step~1 edge for each $x$, into $\Qgl_{i,x,d,e}$. The Step~4 edges of $\Qgl_{i,x,d,e}$ lead to $\Qp$, and their other input is partner B with $\Ftwo$ and $\Fthree$, $\Qgr_{x,d,e,j}$, which lies ahead of the scan; estimating it needs a sum over B's open end $c\in[x,d{-}1]$, so the estimate for $\Qgl$ has $\order(n)$ terms, and the estimate for $\Qg$, summed over its Step~1 edges, $\order((d{-}a)^2)$. As stem B, $\Qg_{x,c,f,j}$ has one Step~2 edge for each $d$, into $\Qgl_{x,d,f,j}$, whose Step~3 edges lead to $\Qgr_{x,d,e,j}$, one for each $e$: $\order(n^2)$ and $\order(n)$ terms again. The Step~4 edges of $\Qgr_{x,d,e,j}$ have as other input partner A with $\Fone$, $\Qgl_{i,x,d,e}$, which lies behind the scan with the same $x$, $d$ and $e$; summed over A's left end $i$ with the prefix, they give the single term above.\end{edited}

The pseudoknot rule $\Qp$ (\cref{eq:Qp}) has eight free indices, so evaluating it as written costs $\order(n^8)$. \DPthree evaluates it in $\order(n^5)$ through auxiliary states that sum out a boundary as soon as every factor that uses it has been multiplied in (\DPthree, Figs.~17--18). This is the principle behind the hook trick of Eisner and Satta~\cite{eisner1999}, and an instance of what Eisner and Blatz~\cite{eisner2007} call folding. \Cref{fig:pkassembly} in the main text carries out the four folding steps. Inside a stem, \DPthree factors the multiloop steps of the gap-state recursion $\Qg$ (\cref{eq:Qg}) in the same way, through $\Qgls$ and $\Qgrs$ (\cref{fig:mlsplit}).

\begin{figure}[p]
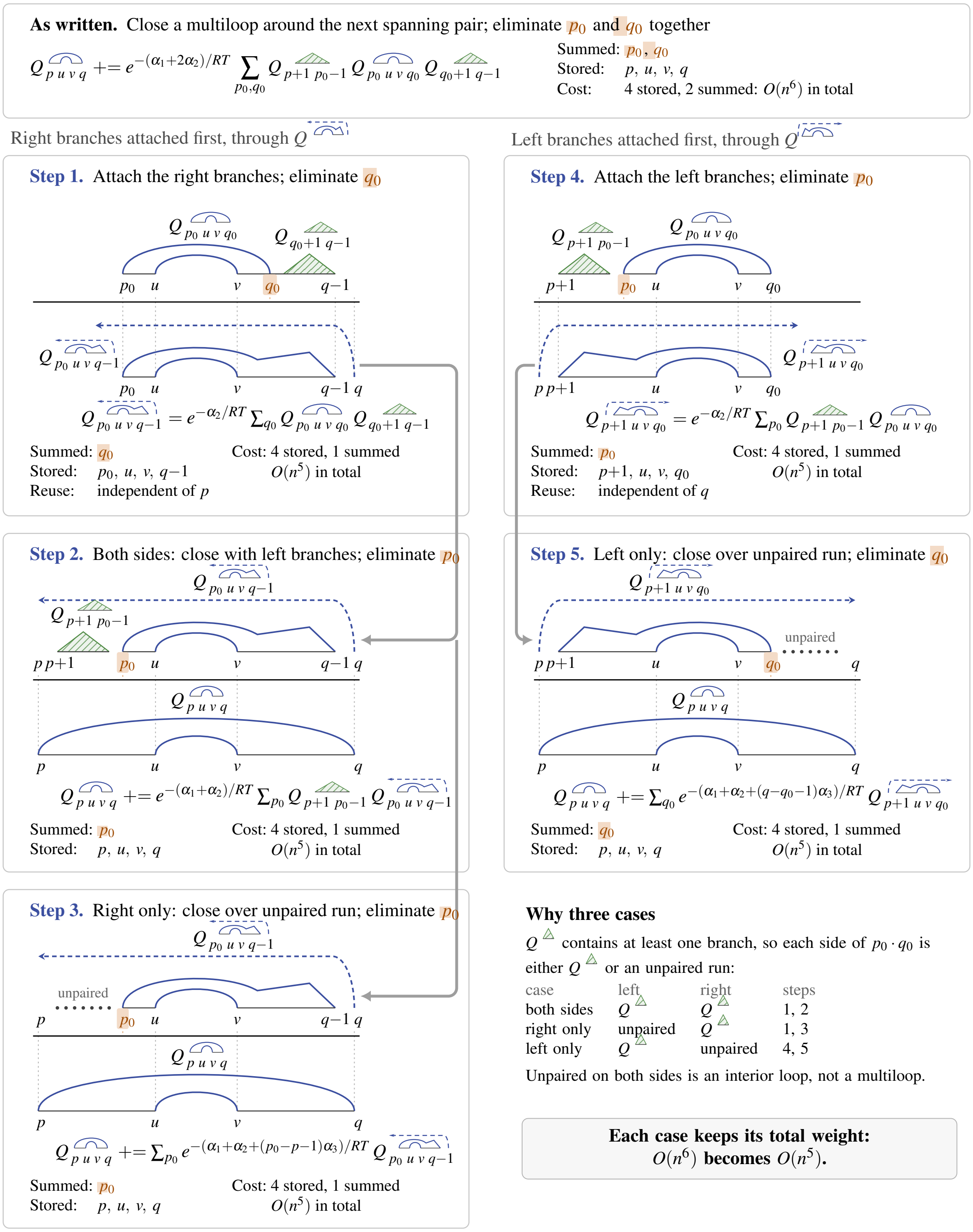

\centering
\input{figs/dp03fig_defs}\providecommand{\Ast}{\mathscr{A}}\providecommand{\Bst}{\mathscr{B}}
\providecommand{\Qring}{Q}\providecommand{\Ql}{Q^{\mathrm{l}}}\providecommand{\Qlx}{Q^{\mathrm{lx}}}
\providecommand{\Qgrseed}{\Qring^{\mathrm{gr}}}\providecommand{\Qgrz}{Q^{\mathrm{grz}}}
\providecommand{\Qpgseed}{\Qring^{\mathrm{pg}}}\providecommand{\Qpg}{Q^{\mathrm{pg}}}
\providecommand{\QtriX}{\accentset{\scriptscriptstyle\triangle}{Q}}\providecommand{\Qtri}{\protect\QtriX}%
\providecommand{\Qtgls}{\Qtri^{\mathrm{gls}}}\providecommand{\Qtgrs}{\Qtri^{\mathrm{grs}}}
\providecommand{\DPfigref}[1]{\ifdefined\hyperref\ref*{#1}\else\ref{#1}\fi}
\providecommand{\DPt}[3]{\node[anchor=base west, inner sep=0pt] at (#1,#2) {#3};}
\providecommand{\DPtc}[4]{\node[anchor=base west, inner sep=0pt, text=#3] at (#1,#2) {#4};}
\providecommand{\DPo}[1]{\hspace{0.7pt}\textcolor{dpelim!80!black}{#1}\hspace{0.7pt}}
\providecommand{\DPinfo}[3]{\node[anchor=base west, inner sep=0pt, font=\footnotesize] at (#1,#2)
  {\begin{tabular}[t]{@{}l@{\ }l@{}}#3\end{tabular}};}
\providecommand{\DPpar}[5]{\node[anchor=base west, inner sep=0pt, text width=#3bp, align=left, #4] at (#1,#2) {#5};}
\providecommand{\DPsp}{\spaceskip=2.6664pt\xspaceskip=4pt\relax}%
\providecommand{\DPspS}{\spaceskip=2.9997pt\xspaceskip=4.5pt\relax}%
\providecommand{\DPinfoS}[3]{\node[anchor=base west, inner sep=0pt, font=\footnotesize] at (#1,#2)
  {\DPsp\begin{tabular}[t]{@{}l@{\ }l@{}}#3\end{tabular}};}
\providecommand{\LKfiller}[3]{\path[fill=black!14, draw=black!55, line width=0.5bp]
  (#1,#3) -- ({(#1+#2)/2},{#3-4.82}) -- (#2,#3) -- cycle;}
\providecommand{\LKarm}[3]{\path[draw=stemB, line width=2.19bp] (#1,#3) -- (#2,#3);}
\providecommand{\LKseedfill}[5]{\LKfiller{#1+2.23}{#2-2.23}{#5}\LKfiller{#3+2.23}{#4-2.23}{#5}\LKarm{#2}{#3}{#5}}
\providecommand{\LKcup}[3]{\pgfmathsetmacro\LKl{#1-2.23}\pgfmathsetmacro\LKr{#2+2.23}%
  \pgfmathsetmacro\LKtl{\LKl-4.03}\pgfmathsetmacro\LKtr{\LKr+4.03}\pgfmathsetmacro\LKk{0.296*(\LKtr-\LKtl)}%
  \path[draw=stemA, line width=0.75bp] (\LKl,#3) -- (\LKtl,{#3-9.30})
    .. controls ({\LKtl+\LKk},{#3-11.68}) and ({\LKtr-\LKk},{#3-11.68}) .. (\LKtr,{#3-9.30}) -- (\LKr,#3);}
\providecommand{\LKcupR}[3]{\pgfmathsetmacro\LKl{#1-2.23}\pgfmathsetmacro\LKr{#2+2.23}%
  \pgfmathsetmacro\LKtr{\LKr+4.03}\pgfmathsetmacro\LKk{0.296*(\LKtr-\LKl)}%
  \path[draw=stemA, line width=0.75bp] (\LKl,#3)
    .. controls (\LKl,{#3-12.4}) and ({\LKtr-\LKk},{#3-11.68}) .. (\LKtr,{#3-9.30}) -- (\LKr,#3);}
\providecommand{\LKcupL}[3]{\pgfmathsetmacro\LKl{#1-2.23}\pgfmathsetmacro\LKr{#2+2.23}%
  \pgfmathsetmacro\LKtl{\LKl-4.03}\pgfmathsetmacro\LKk{0.296*(\LKr-\LKtl)}%
  \path[draw=stemA, line width=0.75bp] (\LKl,#3) -- (\LKtl,{#3-9.30})
    .. controls ({\LKtl+\LKk},{#3-11.68}) and (\LKr,{#3-12.4}) .. (\LKr,#3);}
\providecommand{\DPtx}[3]{\node[anchor=base, inner sep=0pt] at (#1,#2) {#3};}
\providecommand{\DPte}[3]{\node[anchor=base, inner sep=1.1pt, fill=dpelim!22, rounded corners=1pt,
  text=dpelim!80!black] at (#1,#2) {#3};}
\providecommand{\LKplusR}[3]{\pgfmathsetmacro\LKw{#2-#1}\pgfmathsetmacro\LKh{18.7+0.0522*(\LKw-69)}%
  \path[draw=stemA, line width=0.75bp] (#1,#3) .. controls (#1,{#3-\LKh}) and ({#1+0.668*\LKw},{#3-1.12*\LKh})
    .. ({#1+0.83*\LKw},{#3-0.36*\LKh}) -- ({#1+0.94*\LKw},{#3-0.41*\LKh}) -- (#2,#3);}
\providecommand{\LKplusL}[3]{\pgfmathsetmacro\LKw{#2-#1}\pgfmathsetmacro\LKh{18.7+0.0522*(\LKw-69)}%
  \path[draw=stemA, line width=0.75bp] (#2,#3) .. controls (#2,{#3-\LKh}) and ({#2-0.668*\LKw},{#3-1.12*\LKh})
    .. ({#2-0.83*\LKw},{#3-0.36*\LKh}) -- ({#2-0.94*\LKw},{#3-0.41*\LKh}) -- (#1,#3);}
\scalebox{0.93}{\input{figs/fig_ml_split}}
\caption{Factoring the multiloop steps of the gap-state recursion $\Qg$ (\cref{eq:Qg}), in the format of \cref{fig:pkassembly}. Stem $\pair{p}{q}$ closes a multiloop around the next spanning pair $\pair{p_0}{q_0}$ of a gap state with inner pair $\pair{u}{v}$ and gap $[u{+}1,v{-}1]$ (\cref{eq:Qg}). As written (top), the rule sums $p_0$ and $q_0$ together at a cost of $\order(n^6)$, and it has three cases (bottom right). $\Qgrs$ does not depend on $p$ and serves two cases; $\Qgls$ does not depend on $q$. Every row of a step is drawn on one axis. A dashed arc is the closing pair a helper state will receive; dots are an unpaired run; gray arrows carry a stored item to the steps that use it. $\alphaone$, $\alphatwo$, $\alphathree$ are \DPthree's multiloop penalties per loop, per bordering pair and per unpaired base.}
\label{fig:mlsplit}
\end{figure}

\begin{figure}[p]
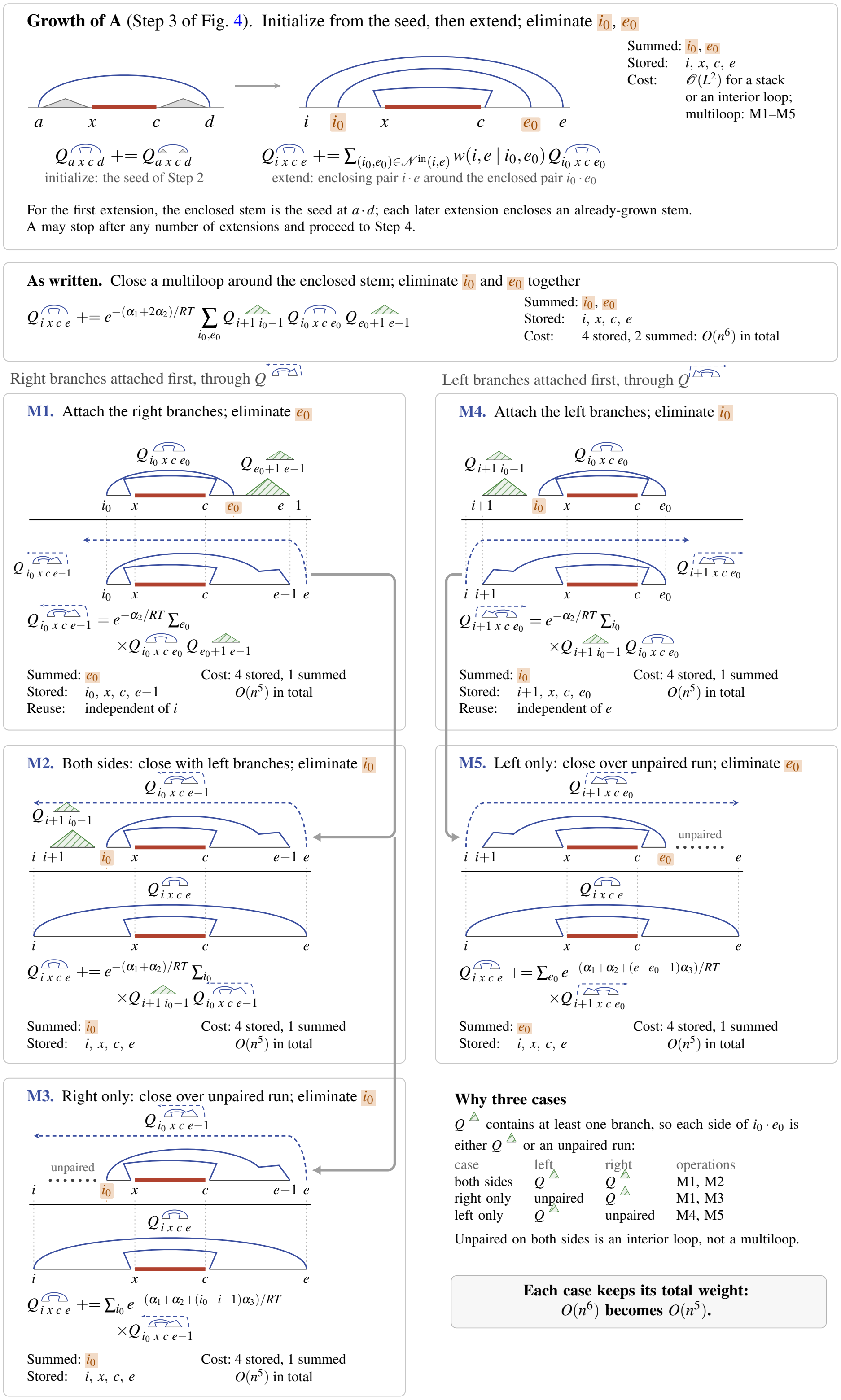

\centering
\input{figs/dp03fig_defs}%
\scalebox{0.745}{\ifdefined\LKcup\else\input{figs/dp03fig_defs}\input{figs/l2rfig_defs}\fi%
\begin{tikzpicture}[x=1bp, y=-1bp]
\path[draw=black!22, line width=0.5bp, rounded corners=3bp] (0.25,3.86) rectangle (467.98,142);
\DPt{13.20}{16.71}{\textbf{Growth of A} (Step~3 of \cref{fig:l2rsteps}).\enspace Initialize from the seed, then extend; eliminate \DPhl{i_0}, \DPhl{e_0}}
\path[draw=black!40, line width=0.4bp] (14,62) -- (124,62);
\path[draw=stemA, line width=0.75bp] (20,62) .. controls (20,38) and (116,38) .. (116,62);
\LKseedfill{20}{50}{86}{116}{62}
\DPtx{20}{73}{$a$}\DPtx{50}{73}{$x$}\DPtx{86}{73}{$c$}\DPtx{116}{73}{$d$}
\DPtx{68}{92}{$\Qgr_{a,x,c,d}\mathrel{+}=\Qgrseed_{a,x,c,d}$}
\node[anchor=base, inner sep=0pt, font=\footnotesize, text=black!65] at (68,104) {initialize: the seed of Step~2};
\path[draw=black!38, line width=1.2bp, -{Stealth[length=4bp]}] (130,50) -- (156,50);
\path[draw=black!40, line width=0.4bp] (166,62) -- (318,62);
\path[draw=stemA, line width=0.75bp] (170,62) .. controls (170,24) and (314,24) .. (314,62);
\path[draw=stemA, line width=0.75bp] (188,62) .. controls (188,33) and (296,33) .. (296,62);
\LKarm{214}{268}{62}\LKcup{214}{268}{62}
\DPtx{170}{73}{$i$}\DPte{188}{73}{$i_0$}\DPtx{214}{73}{$x$}
\DPtx{268}{73}{$c$}\DPte{296}{73}{$e_0$}\DPtx{314}{73}{$e$}
\DPtx{242}{92}{$\Qgr_{i,x,c,e}\mathrel{+}=\sum_{(i_0,e_0)\in\Nin(i,e)}\wloop{i}{e}{i_0}{e_0}\,\Qgr_{i_0,x,c,e_0}$}
\node[anchor=base, inner sep=0pt, font=\footnotesize, text=black!65] at (242,104) {extend: enclosing pair $\pair{i}{e}$ around the enclosed pair $\pair{i_0}{e_0}$};
\DPinfo{350}{30}{Summed: & \DPhl{i_0}, \DPhl{e_0}\\ Stored: & $i,\ x,\ c,\ e$\\ Cost: & $\order(\Lmax^2)$ for a stack\\ & or an interior loop;\\ & multiloop: M1--M5}
\node[anchor=base west, inner sep=0pt, font=\footnotesize, text width=440bp, align=left] at (13.2,122) {For the first extension, the enclosed stem is the seed at $\pair{a}{d}$; each later extension encloses an already-grown stem.\\ A may stop after any number of extensions and proceed to Step~4.};
\end{tikzpicture}%
}\par\vspace{4pt}
\scalebox{0.745}{\input{figs/fig_l2r_ml_split}}%
\caption{Growing stem A (Step~3 of \cref{fig:l2rsteps}). \emph{Top:} the seed of Step~2 initializes the growing table; each extension adds a new enclosing pair $\pair{i}{e}$ around the enclosed stem $\Qgr_{i_0,x,c,e_0}$ (\cref{eq:Qgrgrow}). For the first extension, the enclosed stem is the seed at $\pair{a}{d}$. \emph{Below:} a multiloop extension, factored as in \DPthree (\cref{fig:mlsplit}) by helper operations M1--M5, each summing over one boundary; the helpers are \DPthree's, built on $\Qgr$ instead of $\Qg$ (drawn like $\Qgr$, with both fillers inside). Stem B grows by the same operations in Step~5, keeping its left end $\ge x$.}
\label{fig:l2rmlsplit}
\end{figure}

\paragraph{Single counting in \DPthree.}\label{app:dp03single} A partition function must count each structure once. \DPthree obtains this from McCaskill's rightmost-pair convention~\cite{mccaskill1990}, extended by one kind of element~\cite{dirks2003}. $Q_{k,\ell}$ is split at the rightmost element of the exterior loop of $[k,\ell]$, and every base after it is unpaired. In McCaskill's recursion this element is a pair $\pair{i}{j}$, giving $Q_{k,i-1}\,\Qb_{i,j}$; \DPthree adds a pseudoknot that spans exactly $[i,j]$, giving $Q_{k,i-1}\,\Qp_{i,j}$ \citep[Fig.~12]{dirks2003}. In a given structure the rightmost element is unique: it is the element that contains the last paired base $j$, either a single pair $\pair{i}{j}$ or the pseudoknot that contains $j$, whose first base is $i$. So exactly one term of $Q_{k,\ell}$ produces the structure; in \DPthree's words, a non-redundant $\Qp$ ensures that the algorithm ``never visits a structure twice''~\cite{dirks2003}. $\Qz$ and $\Qm$ use the same convention. Inside $\Qp_{i,j}$, the structure likewise fixes the other six indices of \cref{eq:Qp}; each stem grows by \cref{eq:Qg} one loop at a time, and each filler is split by the same rule. Hence each structure in \DPthree's class has exactly one derivation.

\paragraph{Single counting in \LRDP.}\label{app:l2rsingle} \LRDP changes only how $\Qp$ is evaluated: its six steps compute the same sum, \cref{eq:Qp}, with the same factors, and only regroup it (\cref{tab:elim}). Each of the six inner boundaries is summed out exactly once, at a different step than in \DPthree, and reordering a finite sum neither merges nor duplicates a term. Hence each structure in \DPthree's class has exactly one \LRDP derivation, and without beams \LRDP computes \DPthree's $\Zfn$ in $\order(n^5)$ time and $\order(n^4)$ memory. Pruning only removes derivations, so the beamed $\Zfn$ never exceeds the exact one.

\begin{table}[h]
\centering\small
\caption{Step at which each boundary is summed out (\DPthree: \cref{fig:pkassembly}; \LRDP: \cref{fig:l2rsteps}). Step~3 absorbs the seed boundaries $a,d$ into the growing state; in the no-growth case, they coincide with $i,e$.}
\label{tab:elim}
\setlength{\tabcolsep}{2.2pt}
\begin{tabular}{@{}lcccccc@{}}
\toprule
& $a$ & $x$ & $c$ & $d$ & $e$ & $f$\\
\midrule
\DPthree & Step~1 & Step~4 & Step~2 & Step~4 & Step~4 & Step~3\\
\LRDP & Step~3 & Step~6 & Step~5 & Step~3 & Step~4 & Step~5\\
\bottomrule
\end{tabular}
\end{table}

\section{The left-to-right grammar, step by step}
\label{app:l2r}

\paragraph{What the growing table accumulates (Step~3).} The extension rule of Step~3 (\cref{eq:Qgrgrow}) sums over the immediately enclosed pair $\pair{i_0}{e_0}$, not directly over the original seed $\pair{a}{d}$. Each input entry $\Qgr_{i_0,x,c,e_0}$ already accumulates the contributions of all seeds and growth histories that reach it. The total weight represented by the recurrence is
\begin{equation}\label{eq:Qgrtotal}
\Qgr_{i,x,c,e}=\sum_{a,d}\Qg_{i,a,d,e}\,\Qgrseed_{a,x,c,d}.
\end{equation}
The scan does not evaluate this expression directly, and it does not construct a separate $\Qg$ table.

\paragraph{Deriving the extension rule.} Substituting \DPthree's recurrence for $\Qg_{i,a,d,e}$ (\cref{eq:Qg}) into \cref{eq:Qgrtotal},
\begin{equation*}
\begin{aligned}
\Qgr_{i,x,c,e}&=\textstyle\sum_{a,d}\ind{a=i}\,\ind{d=e}\,\Qgrseed_{a,x,c,d}+\sum_{(i_0,e_0)\in\Nin(i,e)}\wloop{i}{e}{i_0}{e_0}\sum_{a,d}\Qg_{i_0,a,d,e_0}\,\Qgrseed_{a,x,c,d}\\
&=\Qgrseed_{i,x,c,e}+\textstyle\sum_{(i_0,e_0)\in\Nin(i,e)}\wloop{i}{e}{i_0}{e_0}\,\Qgr_{i_0,x,c,e_0}.
\end{aligned}
\end{equation*}
The no-growth term gives the seed contribution; grouping the remaining terms by the immediately enclosed pair, the inner sum is \cref{eq:Qgrtotal} at $\pair{i_0}{e_0}$. Together these are the initialization and the extension rule of Step~3. The seed term is zero wherever no seed is valid. This derivation establishes the identity for the unpruned computation; with beams, the updates use the retained states.

\paragraph{Multiloop extensions.} A stack or an interior loop sums over $\order(\Lmax^2)$ enclosed pairs $\pair{i_0}{e_0}\in\Nin(i,e)$ per state. A multiloop extension, as written, sums over $i_0$ and $e_0$ together, an $\order(n^6)$ rule over $i,x,c,e,i_0,e_0$. As in \DPthree (\cref{fig:mlsplit}), helper states built on $\Qgr$ instead of $\Qg$ ($\Qtgls$ and $\Qtgrs$, drawn like $\Qgr$ with both fillers inside) factor it into operations that each sum over one boundary, M1--M5 in \cref{fig:l2rmlsplit}, at $\order(n^5)$ in total, the cost of \DPthree's own gap-state recursion; only kept $\Qgr$ states are extended. Stem B grows by the same operations in Step~5, keeping its left end $\ge x$; unrolled in the same way, Steps~5 and~6 give $\Qp_{i,j}=\sum_{x,c,f}\Qg_{x,c,f,j}\,e^{-\betatwo/\RT}\,\Qgrz_{i,x,c,f-1}$: B's weight is \DPthree's $\Qg_{x,c,f,j}$.

\paragraph{Exact outside weights of the \LRDP states.}\label{app:keysexact} \begin{edited}For a pseudoknot in the exterior loop, with $\Lambda$, $\Vex$ and $\KLex$ of \cref{eq:Vex,eq:KLex}, the four stem states of \cref{fig:l2rsteps} have
\begin{equation}\label{eq:keysall}
\begin{aligned}
\Zout(\Qgrseed_{a,x,c,d})&=\omegaone\textstyle\sum_{e\ge d}\Lambda(a,d)[e]\,\KLex(e;x,c), &&\text{seed of A (Step~2)},\\
\Zout(\Qgr_{i_0,x,c,e_0})&=\omegaone\textstyle\sum_{e\ge e_0}\Lambda(i_0,e_0)[e]\,\KLex(e;x,c), &&\text{grown A (Step~3)},\\
\Zout(\Qgrz_{i,x,c,t})&=\omegaone\,\Cpref(i{-}1)\,\KLex(t;x,c), &&\text{closed A, reading }\Fthree\text{ (Step~4)},\\
\Zout(\Qpg_{i,x,p,q})&=\omegaone\,\Cpref(i{-}1)\,\Vex[p][q](x), &&\text{growing B, from its seed (Step~5)}.
\end{aligned}
\end{equation}
$\Lambda(i_0,e_0)[e]$ includes the term with no further growth ($i=i_0$, $e=e_0$), in which A stops at $\pair{i_0}{e_0}$. In $\Qgrz$, $t$ is the scan position and the rest of $\Fthree$ is a filler on $[t{+}1,f{-}1]$, so $\KLex$ is taken at $t$. Expanding $\Vex$ by \cref{eq:Vex} as a recursion over $\Nout(c,f)$, stopped when B's left end reaches $x$, gives the form that Pass~2 computes (\cref{eq:V}).\end{edited}

\begin{edited}\paragraph{Outside estimates of the scan states.}\label{app:keys} This gives the estimates of the outside weights in \cref{eq:keysall}. What remains to be estimated depends on how far the construction has progressed (\cref{fig:l2rsteps}). \emph{Seed and grown stem A, $\Qgrseed$ and $\Qgr$ (Steps~2--3).} A can still grow: for each final right end $e$, its estimate combines A's growth and the prefix before it with the completion by stem B, the third filler $\Fthree$ and the suffix. We combine A's estimated growth weight $\hG$ (next paragraph) with the prefix weight $\Cpref$ and sum over A's final left end $i$: $\hLam(i_0,e_0)[e]=\sum_{i\le i_0}\Cpref(i{-}1)\,\hG(i_0,e_0)[i,e]$ estimates A's growth from its current outer pair $\pair{i_0}{e_0}$ to the final right end $e$, prefix included. \emph{Closed stem A with a partially read $\Fthree$, $\Qgrz$ (Step~4).} The state already holds A's growth; only the rest of $\Fthree$, stem B and the suffix remain, which is the completion weight $\KL$ (\cref{eq:KL}) at the scan position $t$. \emph{Growing stem B, $\Qpg$ (Step~5).} The state holds stem A and all three fillers; only B's growth to the reserved $x$ and the suffix remain, which is one lookup in the partner table $\hV$ (\cref{eq:V}). In the other cases we also multiply by the prefix weight $\Cpref(i{-}1)$; for growing stem A, the prefix is already included in $\hLam$. In \cref{eq:keysall}, $t$ is any scan position, not only the $f{-}1$ at which B is seeded. A seed has the estimate of its grown state. With exact tables and the pseudoknot in the exterior loop, each estimate is exact. $\KL(t;x,c)$ sums over the inner ends $f$ of the partner entries $\hV[c][f](x)$ that Pass~2 kept for the arm $(x,c)$. \end{edited}

\begin{edited}\paragraph{A's side, on demand.}\label[appendix]{sec:growth} Only growing stem A needs the growth estimate $\hLam$ (\cref{eq:keys}). Rather than computing it for every possible stem, Pass~3 evaluates it when a candidate state needs a score, and caches the result for reuse. Each growth step of the stem-growth recurrence (\cref{eq:Qg}) has a weight independent of A's final left end $i$, so $\hLam$ follows the same outward, one-loop-at-a-time rule:
\begin{blackmath}
\begin{equation}
\hLam(i_0,e_0)[e]=\ind{e=e_0}\,\Cpref(i_0{-}1)+\sum_{(i',e')\in\Nout(i_0,e_0)}\wloop{i'}{e'}{i_0}{e_0}\;\hLam(i',e')[e].\label{eq:lam}
\end{equation}
\end{blackmath}
The first term stops growth at the current pair and supplies the prefix weight; the second adds one outer loop and continues from the new pair. Pass~3 keeps $K=32$ right ends $e$ per state.\end{edited}

\paragraph{Why Pass~3, not Pass~2, builds A's side.} A's growth has its final left end $i$ before the scored stem, so the scan can rank it with the true prefix $\Cpref$, pseudoknots included, rather than Pass~2's nested $\Cpre$, and it needs growth only for the stems it scores (\cref{sec:eff}); stem B cannot yet be built, because its right arm lies beyond the scan, so Pass~2 supplies the partner estimate $\hV$.

\paragraph{Incremental spawn.} Creating a seed $\Qgrseed$ for every arm at every inner pair would create $\order(n^4)$ candidates before any pruning: 623 million at 486\nt. Following \LinearPartition, we instead create every state from a beamed state, so the arm is built in scan order: the first filler $\Qz_{a+1,x-1}$ and $\Qlx$ are themselves beamed tables (Steps~1--2), ranked by the weight of completing their open filler, and a seed is created only from a $\Qlx$ that survived. Unbeamed, Steps~1--2 reproduce every seed exactly. Every rule of the scan makes a state that ends at the scan position $t$ in one of three ways: it extends one kept state that ends at $t{-}1$ by one base (an unpaired base of a filler, or one more base of B's arm); it joins two kept states that meet at one position (an item that ends at $t$ with a state that ends where the item starts, or, for a seed, a filler $\Qz_{c+1,t-1}$ with the $\Qlx_{a,x,c}$ that ends where the filler starts); or it grows a kept stem by the stem-growth recurrence (\cref{eq:Qg}), over $\order(\Lmax^2)$ inner pairs or one multiloop helper. No rule extends a state to all later positions at once, and no rule reads all earlier positions. Each table keeps at most $\btwo$ or $\bfour$ states per position, so each position combines $\order(\btwo\,\bfour)$ pairs of kept states, and the scan is linear in $n$.

\section{Pass 2 details}
\label{app:pass2}

\paragraph{Ranking keys of Pass~2.} $\Qzpre$ is built left to right and keeps, for each right end $f{-}1$, the $\btwo$ start points $e{+}1$ with the largest $\Cpre(e)\,\Qzpre(e{+}1,f{-}1)$. $\hV$ keeps, in each map $\hV[c][f](\cdot)$, whose entries all leave the gap $c{+}1..f{-}1$ open, the $\bV$ ends $x$ with the largest $\Cpre(x{-}1)\,\hV[c][f](x)$; it is built in one outside pass, right to left over the states Pass~1 keeps, as \LinearPartition's outside pass is. Both keys multiply each entry by Pass~1's nested prefix up to its left end, so the entries they compare cover the same part of the sequence.

\paragraph{A's side in Pass~3.}\label{app:aside} Building A's side $\hLam$ in Pass~3 rather than in Pass~2 matters for both accuracy and cost. Ranking growth by Pass~2's nested prefix $\Cpre$ undervalues pseudoknot stems: on tmRNA \texttt{Ther.ther} it drops both stems of the pseudoknot and loses 12.8 in $\log\Zfn$. On 16 tmRNA and RNase P sequences, growth on demand uses 2.6 times less CPU time than growth in Pass~2 (2{,}183\,s vs.\ 5{,}760\,s) at the same accuracy (mean $|\Delta\log\Zfn|$ 0.302 vs.\ 0.304). %

Unbeamed, Pass~2's outside pass is exact: on RNase P 1--150, $\sum_x\Cpre(x{-}1)\,\hV[c][f](x)$ reproduces Pass~1's outside weight of every pair $\pair{c}{f}$ within a relative error of $8\times10^{-13}$.

\section{Structure accuracy metrics}
\label{app:metrics}
For one sequence, let $P$ be the set of predicted pairs and $\GT$ the set of reference pairs; a predicted pair is correct only if both its indices match a reference pair. Two pairs $(i,j)$ and $(k,\ell)$ \emph{cross} if $i<k<j<\ell$ or $k<i<\ell<j$, and for $S\in\{P,\GT\}$, $S_{\mathrm{cross}}=\{s\in S: s \text{ crosses another pair of } S\}$. In every case, $\mathrm{F1}=2\,\mathrm{Prec}\,\mathrm{Rec}/(\mathrm{Prec}+\mathrm{Rec})$.

\paragraph{All pairs.} $\mathrm{Prec}=|P\cap\GT|/|P|$ and $\mathrm{Rec}=|P\cap\GT|/|\GT|$.

\paragraph{Crossing pairs.} A predicted pair counts only if it is in $\GT$ and is crossing in both structures:
\begin{equation*}
\mathrm{Prec}^{\mathrm{cross}}=\frac{|P_{\mathrm{cross}}\cap\GT_{\mathrm{cross}}|}{|P_{\mathrm{cross}}|},\qquad
\mathrm{Rec}^{\mathrm{cross}}=\frac{|P_{\mathrm{cross}}\cap\GT_{\mathrm{cross}}|}{|\GT_{\mathrm{cross}}|}.
\end{equation*}

\section{The Rous sarcoma virus element: all methods}
\label{app:rsv}
\Cref{fig:rsvall} compares all methods on the 114\nt pseudoknot of the \emph{gag-pol} element of \cref{sec:rsv} (positions 2489--2602, the structure 9DID~\cite{jones2025rsv} without the slippery site and spacer). \LN and \HotKnots~2.0 form the full pseudoknot of helices A--B and D: each predicts 8 of D's 9 pairs (for \LN, with probability at least 0.5). \IPknot (RNAfold) predicts 5 of D's pairs but neither pair of B. LinearPartition-V and LinearPartition-C with \ThreshKnot, CParty with \ThreshKnot, and \IPknot (LP-C) predict no pair of D. Every method finds all 9 pairs of helix A, and none predicts helix C, which lies outside \DPthree's class. The 1-best structure of \LN is the most accurate (F1 0.82; \HotKnots 0.79, \IPknot (RNAfold) 0.78, the others 0.50 to 0.70).

\begin{figure}[h]
\centering
\includegraphics[width=\linewidth]{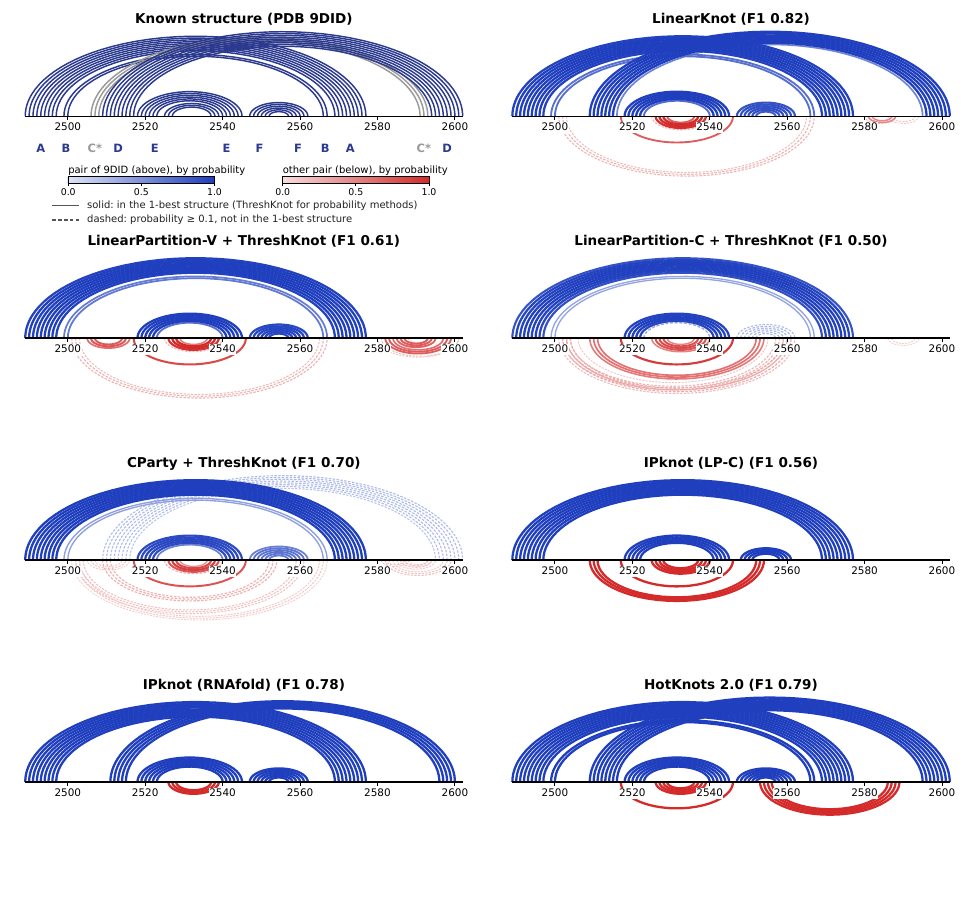}
\caption{The \emph{gag-pol} frameshift element of Rous sarcoma virus (NC\_001407.1, positions 2489--2602, the pseudoknot only) for every method, in the format of \cref{fig:rsv}. Top left: the structure 9DID; helix C (gray) lies outside \DPthree's class. Methods with a partition function (\LN at the default beams, LinearPartition-V, LinearPartition-C and CParty, the last three as used by \ThreshKnot) are drawn as pair probabilities at least 0.1; \IPknot and \HotKnots~2.0 are drawn as their single predicted structure. Pairs of 9DID are drawn above the axis (blue), others below (red); darker is more probable. Solid: pairs in the method's 1-best structure (\ThreshKnot with $\theta=0.3$ for the partition-function methods; all pairs for \IPknot and \HotKnots); dashed: pairs with probability at least 0.1 that are not in the 1-best structure. Panel titles give the F1 of the 1-best structure against 9DID. CParty and \HotKnots use \DPnine parameters~\cite{andronescu2010}.}
\label{fig:rsvall}
\end{figure}

\fi

\end{document}